\documentclass[aps,twocolumn,superscriptaddress,floatfix,longbibliography]{revtex4-1}

\usepackage{lipsum}
\usepackage{graphicx}
\usepackage{amsmath,amssymb}
\usepackage{braket}
\usepackage{mathtools}
\usepackage{hyperref}

\usepackage{xcolor}
\usepackage[normalem]{ulem}

\newcommand{\Z}{\mathbb{Z}}
\newcommand{\st}[1]{\left\{ #1 \right\}}%%adjustable-height set notation

\begin{document}

\title{Gapless Symmetry Protected Topological Order}

%random order of names for now
\author{Thomas Scaffidi}
\email[]{thomas.scaffidi@berkeley.edu}
\affiliation{Department of Physics, University of California, Berkeley, CA 94720, USA}
\author{Daniel E. Parker}
\email[]{daniel\_parker@berkeley.edu}
\affiliation{Department of Physics, University of California, Berkeley, CA 94720, USA}
\author{Romain Vasseur}
\email[]{rvasseur@umass.edu}
\affiliation{Department of Physics, University of California, Berkeley, CA 94720, USA}
\affiliation{Materials Sciences Division, Lawrence Berkeley National Laboratory, Berkeley, CA 94720, USA}
\affiliation{Department of Physics, University of Massachusetts, Amherst, MA 01003, USA}

\date{\today}

\begin{abstract}

	We introduce exactly solvable gapless quantum systems in $d$ dimensions that support symmetry protected topological (SPT) edge modes. Our construction leads to long-range entangled, critical points or phases that can be interpreted as critical condensates of domain walls ``decorated'' with dimension $(d-1)$ SPT systems. Using a combination of field theory and exact lattice results, we argue that such gapless SPT systems have symmetry-protected topological edge modes that can be either gapless or symmetry-broken, leading to unusual surface critical properties. Despite the absence of a bulk gap, these edge modes are robust against arbitrary symmetry-preserving local perturbations near the edges. In two dimensions, we construct wavefunctions that can also be interpreted as unusual quantum critical points with diffusive scaling in the bulk but ballistic edge dynamics.

\end{abstract}
\maketitle

\section{Introduction}
%%SPTs
%%Gapless topological -> Weyl
%%RK criticality

An overarching goal of condensed matter physics is to identify and classify new phases of matter. Since probing a system amounts to perturbing it and measuring how it reacts, understanding the physics of a phase reduces to the problem of identifying the low-lying excitations that perturbations can create.
A natural dichotomy is to distinguish gapless phases, which possess excitations arbitrarily close to the ground state, from gapped ones, which have a finite spectral gap in the thermodynamic limit. Naively, this would suggest gapped systems are featureless at low energy.

Discoveries in recent decades have shown the story is more subtle, as a large class of gapped phases can host gapless excitations localized to edges and defects. Such excitations are protected by a combination of symmetries and the topological properties of the bulk system. These topological phases include long-range entangled systems~\cite{PhysRevLett.96.110404,PhysRevLett.96.110405} with intrinsic topological order and bulk anyonic excitations, such as quantum Hall states or spin liquids~\cite{wen2004quantum}. They can be further enriched by symmetries~\cite{PhysRevB.65.165113,PhysRevLett.105.246809,PhysRevB.83.195139,PhysRevB.86.115131,PhysRevB.87.155115,YuanMing2013, WenEnriched2013}.  Following the theoretical prediction and subsequent experimental discovery of topological insulators and superconductors \cite{PhysRevLett.95.226801,Bernevig1757,Konig766,PhysRevLett.98.106803,PhysRevB.75.121306,PhysRevB.79.195322,Hsieh:2008aa,Hasan2010,Rasche2013,story2012}, attention has turned to short-range entangled phases with topological edge modes protected by symmetry~\cite{PhysRevB.80.155131,PhysRevB.84.235128,PhysRevB.83.075102,PhysRevB.83.075103,Chen2011b,Pollmann2012,YuanMing2012,Levin2012,Chen1604,Chen2011}. These symmetry protected topological (SPT) phases may be realized in strongly interacting systems, like the experimentally accessible Haldane phase in quantum spin chains~\cite{Buyers1986}. This shift in paradigm from band topology analysis of non-interacting Hamiltonians~\cite{doi:10.1063/1.3149495,PhysRevB.78.195125} to strongly correlated systems led to the development of non-perturbative techniques, resulting in an essentially exhaustive classification of gapped bosonic~\cite{Chen1604,Chen2011,YuanMing2012,PhysRevB.91.134404} and, to some extent, of fermionic SPT phases~\cite{PhysRevB.81.134509,PhysRevB.83.075103,PhysRevB.89.201113,Wang629,PhysRevB.90.115141,Gu2015}. All these phases enjoy a bulk spectral gap and indeed this gap often plays a crucial role in understanding topological phases.

Must systems have a bulk gap to possess the properties of topological phases? 
%Perhaps the bulk gap is unnecessary, and simply makes it easier to note the presence of, e.g., exotic edge effects, as they are the only low-energy features of systems with bulk gaps.
Given the prevalence of gapless systems in nature, it is possible that many of the features ascribed to gapped topological systems are ``hidden'' around their edges~\cite{bonderson2013quasi}. As an example of a step in this direction, it was recently argued that topological phases can survive in non-equilibrium, highly-excited states where there is no notion of a gap~\cite{PhysRevB.88.014206,1742-5468-2013-09-P09005,Bahri:2015aa}. 
	In the less exotic realm of equilibrium physics at low temperature, Weyl and Dirac semi-metals with topologically-protected Fermi arc surface states~\cite{PhysRevB.83.205101} are gapless systems with topological properties that have been experimentally confirmed in several materials~\cite{PhysRevX.5.031013,Lu622,Xuaaa9297}.
	Other examples related to free-fermionic systems include the A phase of superfluid $^3$He~\cite{volovik2003universe}, power-law superconducting chains~\cite{PhysRevB.84.195436,PhysRevB.84.144509,PhysRevLett.114.100401,PhysRevB.91.235309} and recent proposals for gapless topological insulators~\cite{PhysRevLett.114.136801} and superconductors~\cite{1367-2630-15-6-065001,PhysRevB.92.045128}.
		
		Examples of gapless topological systems~\cite{bonderson2013quasi} are, for the most part, restricted to non-interacting systems. 
		Some exceptions include topological Mott insulators~\cite{Pesin:2010aa}, topological Luttinger liquids~\cite{2017arXiv170402997J,1367-2630-16-9-093040}, gapless spin liquids~\cite{PhysRevB.65.165113,0034-4885-80-1-016502}, the Gaffnian quantum Hall state~\cite{PhysRevB.75.075317}, and the composite Fermi liquids in the half-filled Landau level~\cite{PhysRevB.47.7312,PhysRevX.5.031027}. 
		%This is already sufficient evidence to answer the above question in the negative --- gapless systems systems may indeed possess topological properties. 
		However, the precise topological nature --- and edge properties --- of many of these systems remains controversial.

		In this work, we present a general construction of strongly interacting, long-range entangled, quantum systems that are gapless in the bulk with topological edge modes protected by symmetry. These gapless symmetry protected topological states of matter are generated via a systematic procedure that employs standard tools of gapped SPT phases, making their topological properties transparent. For concision, we refer to them as ``gapless SPTs'' (gSPTs). Just as normal SPTs can be thought of as ``twisted'' paramagnets, gapless SPTs can be obtained by twisting ordinary quantum critical points or critical phases.
%\Romain{since they can be obtained by ``twisting'' ordinary quantum critical points or critical phases just like SPTs can be thought of as twisted paramagnets}.
		Some examples of gapless SPTs may be produced starting from an SPT and tuning a subset of the degrees of freedom to criticality. 
		
%		Aside from the lack of a bulk gap, and the resulting long-range entanglement, gSPTs share the other properties of SPTs. 
%		They lie outside existing classifications of symmetry-protected topological order.

		In Section II, we outline the general construction based on the decorated domain wall picture of gapped SPT phases~\cite{AshvinDecorated}. This yields many examples, but we focus on several with the virtue of being exactly solvable: a topological critical Ising chain and a topological Luttinger liquid phase in one dimension (Section III), and a topological gapless spin liquid in two dimensions (Section IV). In all cases we start with the parent Hamiltonian, find the exact ground state wavefunction, and demonstrate the presence of topologically protected edge modes that must be either gapless or symmetry-broken. Despite the absence of a bulk gap, the topological edge modes in such gSPT systems are robust to arbitrary symmetry-preserving boundary perturbations and require no fine-tuning beyond closing the bulk gap. In particular, our general construction can be applied to both quantum critical points and gapless phases.

The topological edge modes of gSPTs can be interpreted as giving rise to exotic surface criticality \cite{PhysRevLett.118.087201}.
%		\sout{The topological edge modes in gapless SPTs have rich physical consequences. One can interpret gSPTs as quantum critical points with exotic surface critical properties} . 
Below we show this can take the form of anomalous edge magnetization, or the appearance of ballistic dynamics at the edge of a diffusive system. Our construction therefore yields a host of gapless systems that blend the physics of quantum critical and topological systems.

\begin{figure}
	\includegraphics{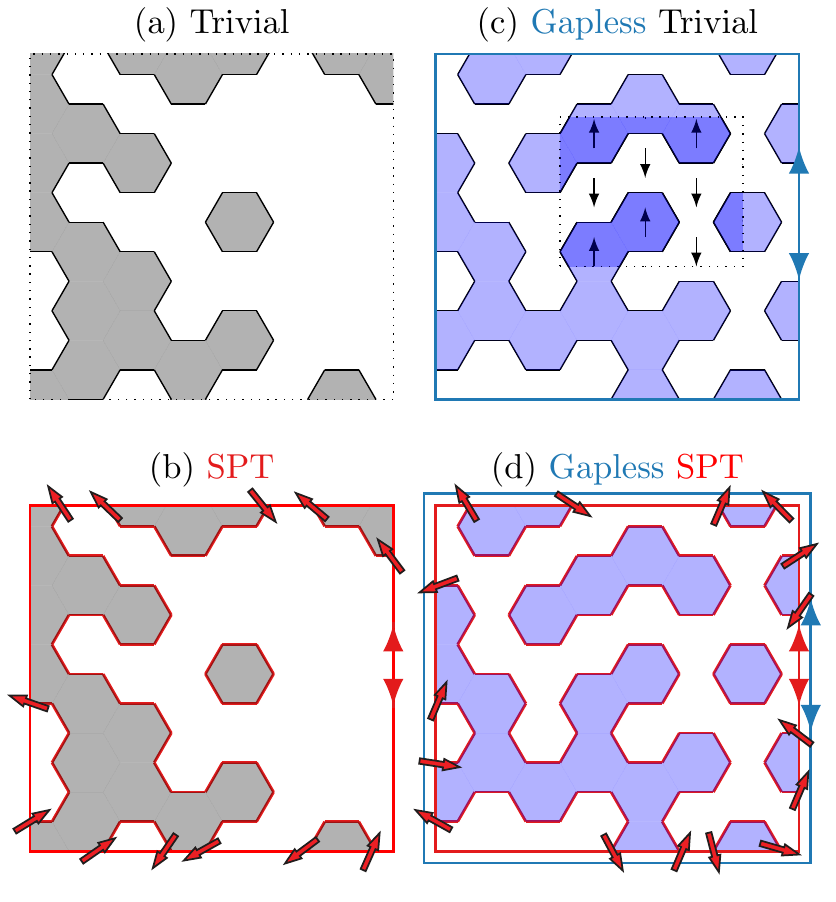}
	\caption{Representative states of each order in the 2D example. \textit{(a) Trivial:} Paramagnetic spins on a triangular lattice with fluctuating domain walls. \textit{(b) SPT:} Decorating the domain walls gives an SPT with a $c=1$ edge mode. \textit{(c) Gapless Trivial:} Tuning the domain walls to criticality by restricting them to fully-packed loop configurations (defined below) closes the gap and gives a $c=1$ edge mode. \textit{(d) Gapless SPT:} Doing both yields a gapless SPT with $c=1+1=2$ edge modes. } 
		\label{fig:cartoon}
\end{figure}

gene

\section{General Construction}

Consider a bosonic system in $d$ dimensions composed of $\sigma$ and $\tau$ degrees of freedom and symmetry group $G_\sigma \times G_\tau$ with $G_\sigma=\Z_2$. Our construction starts from the decorated domain wall picture of SPTs~\cite{AshvinDecorated}. In this picture, a ``trivial'' disordered phase (``trivial'' paramagnets, Fig.~\ref{fig:cartoon} (a)) is thought of as a gapped condensate of domain walls.  Non-trivial SPT phases (``topological'' paramagnets, Fig.~\ref{fig:cartoon} (b)) are produced by ``decorating'' the domain walls of $G_\sigma=\Z_2$ with $(d-1)$ dimensional SPT phases protected by the symmetry $G_\tau$. The protected edge modes appear naturally: domain walls that end at a boundary carry the topologically protected edge mode of the lower-dimensional SPT.

%Our construction starts from the decorated domain wall picture of SPTs~\cite{AshvinDecorated}. In this picture, a ``trivial'' disordered phase (``trivial'' paramagnets, Fig.~\ref{fig:cartoon} (a)) is thought of as a gapped condensate of domain walls. Non-trivial SPT phases (``topological'' paramagnets, Fig.~\ref{fig:cartoon} (b)) are produced by ``decorating'' the domain walls with lower-dimensional SPT phases. The protected edge modes appear naturally: domain walls that end at a boundary carry the topologically protected edge mode of the lower-dimensional SPT. In the following we focus on bosonic systems in $d$ dimensions with symmetry group $G_\sigma \times G_\tau$, where the domain walls of the $G_\sigma = \Z_2$ symmetry are decorated by $(d-1)$ dimensional SPT states protected by the symmetry group $G_\tau$. 

To make a gapless system, we tune the domain wall condensate to criticality (i.e. tune the underlying $\sigma$ degrees of freedom to criticality). When the domain walls are not decorated, (the ``gTrivial'' case, Fig.~\ref{fig:cartoon} (c)), this typically tunes the system to an ordinary quantum critical point. For example, in 1D, one can consider the domain walls of a critical Ising chain and, in 2D, one can use the domain walls of an Ising frustrated antiferromagnet. Generically, there is nothing protected about the edge of such gTrivial systems: they may or may not have additional gapless modes at their boundaries. 

The crucial step is that one may decorate the gTrivial system with lower-dimensional SPT systems. This leads to a topologically distinct gapless state (called ``gSPT'', Fig.~\ref{fig:cartoon} (d)) which, in analogy to the gapped case, has the same properties as gTrivial in the bulk, but completely different edge physics. Topologically protected edge modes appear in gSPT that can be gapped out only at the price of breaking the symmetry at the edge (either spontaneously or explicitly). In short, starting from a gapped SPT, one can generate a gapless SPT by making the domain wall condensate critical while keeping the same domain wall decoration. 

The resulting gSPT systems are tuned to criticality in the bulk, while the edge modes are robust against symmetry-preserving perturbations acting near the edge. Even though some of the examples we treat in this work correspond to critical points, as opposed to gapless phases, this is by no means a limitation of our construction. (To be clear, gapless SPTs are \textit{not} ``symmetry protected gapless phases'' \cite{2016arXiv160605652L,PhysRevLett.118.021601,PhysRevB.96.125104}  --- the gaplessness of the bulk theory is not protected by symmetry.) As we will show explicitly below, the same construction of applying the SPT decoration can be performed in gapless phases, such as Luttinger liquids in 1D \cite{2017arXiv170402997J} or gapless spin liquids in 3D \cite{2015arXiv151101505S}, to obtain gapless SPT phases. More generally, gSPTs are as stable as their underlying gTrivial states before applying the decoration.
In particular, gSPTs have exactly the same spectrum as their parent gTrivial systems on closed manifolds, since they are related by a local unitary transformation.

\section{One Dimension}

\begin{figure}
	\includegraphics[width=3.375in]{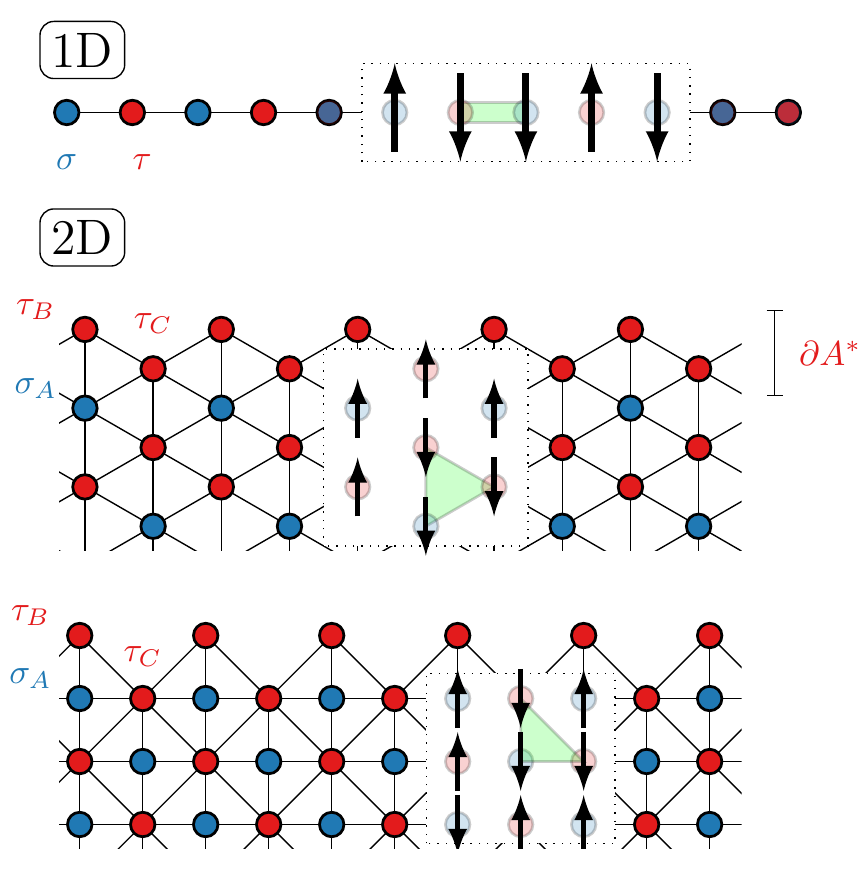}
	\caption{Lattices used throughout this paper to construct SPT and gSPT wavefunctions in one (Top) and two dimensions (Bottom: Triangular and Union Jack lattices). The control-$Z$ twist operator used to obtain non-trivial SPT order gives a factor of $(-1)$  to links with two down spins in the 1D case and triangles with three down spins in the 2D case, as exemplified by the green shading.}
	\label{fig:lattices}
\end{figure}

This section provides a first example of a gapless SPT in one dimension, combining the features of a well-understood 1D gapped SPT and of the critical Ising model. Starting from a gapped SPT with $\Z_2\times \Z_2$ symmetry, we bring one of the spin species to criticality and argue in this exactly solvable limit that this gapless system has topological edge modes. Going beyond this exactly solvable limit, we numerically demonstrate the robustness of these symmetry-protected topological edge modes against arbitrary symmetry-preserving perturbations. 

\subsection{Gapped $\Z_2\times \Z_2$ SPT}

To set the notation, we first recall the construction of a gapped SPT with $\Z_2\times \Z_2$ symmetry in one dimension~\cite{Chen2011,AshvinDecorated,Bahri:2015aa,PhysRevB.93.155131}, which is closely related to the experimentally observable Haldane phase~\cite{PhysRevLett.61.1029,Kennedy1992,Buyers1986}. Consider a spin-$1/2$ chain with two alternating spin species: $\sigma$ (on sites $i$) and $\tau$ (on sites $i+\tfrac{1}{2}$), as shown in Figure \ref{fig:lattices}. We impose an inviolable $\Z_2 \times \Z_2 = G_\sigma \times G_\tau$ global symmetry generated by ${\cal C}_\sigma = \prod_i \sigma_i^x$ and ${\cal C}_\tau = \prod_i \tau_{i+\frac{1}{2}}^x$. In 1D, $\Z_2 \times \Z_2$ is the minimal symmetry required to have a non-trivial SPT.

A trivial paramagnetic phase is obtained with the zero-correlation length Hamiltonian 
\begin{equation}
	H_\text{Trivial} =  - \sum_i \sigma_i^x  +\tau_{i-\frac{1}{2}}^x,
	\label{eq:PM_1d}
\end{equation}
with ground state wavefunction
\begin{equation}
	\ket{\Psi_\text{Trivial}} =  \sum_{\st{\sigma^z},\st{\tau^z}} \ket{ \sigma^z,\tau^z},
	\label{eq:PM_1d_wf}
\end{equation}
where the sum runs over all $\sigma^z$ and $\tau^z$ configurations. This can be thought of as a gapped phase where domain walls have ``proliferated''. 

%	In the $Z$-basis $\CZ = \operatorname{diag}(1,1,1,-1)$, i.e. it gives a phase factor of $(-1)$ only if both spins point down --- see Figure \ref{fig:lattices}.}
%, whose action in the $Z$ basis is $(-1)$ if the two spins are down, and $+1$ otherwise, i.e. $\CZ = (-1)^{\delta_{\downarrow\downarrow}}$ --- see Figure \ref{fig:lattices} \Romain{TODO: add expression in terms of Pauli matrices, get rid of CZ}.

%A \Dan{solvable point in a} non-trivial SPT phase can then be made by ``twisting'' or ``decorating'' this Hamiltonian by a local unitary operator $U_\text{1D}$ \cite{Chen2011}.
%\Dan{Define:
%	\begin{equation}
%		U_\text{1D} = \prod_i (-1)^{P_i P_{i+\frac{1}{2}}} (-1)^{P_{i+\frac{1}{2}} P_{i+1}}
%		%\prod_i (-1)^{\frac{1}{4}\left( 1+\sigma^z_i\right) \left( 1+\tau^z_{i+1/2} \right)}
%		%(-1)^{\frac{1}{4}\left( 1+\tau^z_{i+1/2}\right) \left( 1+\sigma^z_{i+1} \right)}
%		\label{eq:U_1D_defn}
%	\end{equation}
%	where $P_i = \frac{1}{2}\left( 1- \sigma^z_i \right)$ is a projector onto $\ket{\downarrow}$. More physically, this gives a factor of $(-1)$ for each adjacent pair of spin downs in the $z$-basis. (Also called the control-$Z$ operator.) Since the action of $U_\text{1D}$ is just giving phase factors in the $z$-basis, we employ the notation $U_\text{1D} =  e^{i \theta_\text{1D}\left( \sigma^z,\tau^z \right)}$.}

A exactly solvable example of a non-trivial SPT phase can then be made by ``twisting'' or ``decorating'' this Hamiltonian by a local unitary operator $U_\text{1D}$~\cite{Chen2011}. Define 
\begin{equation}
U_\text{1D}=\prod_i \rm{CZ}_{i-1,i-1/2} \rm{CZ}_{i,i-1/2},
\label{eqUnitaryTwist1D}
\end{equation}
 where $\rm{CZ}_{ij} = (-1)^{\delta_{\downarrow\downarrow}}$ is the control-$Z$ two-qbit operator  with $\delta_{\downarrow\downarrow} = \tfrac{1}{4} (1-\sigma_i^z)(1-\tau_j^z)$, which gives a $(-1)$ if the two spins are down and a $+1$ otherwise --- see Figure \ref{fig:lattices}. Alternatively, $U_\text{1D}$ can be thought of as attaching charges of one $\Z_2$ symmetry to domain walls of the other $\Z_2$ symmetry~\cite{PhysRevB.91.195117}. Under periodic boundary conditions, this unitary transformation commutes with the $\Z_2 \times \Z_2$ symmetry.  Explicitly, the non-trivial SPT Hamiltonian $H_\text{SPT} = U_\text{1D} H_\text{Trivial}  U_\text{1D}$ reads
\begin{equation}
	H_\text{SPT} = - \sum_i \tau_{i-\frac{1}{2}}^z \sigma_i^x \tau_{i+\frac{1}{2}}^z +\sigma_{i-1}^z \tau_{i-\frac{1}{2}}^x \sigma_{i}^z,
	\label{eq:SPT_1d}
\end{equation}
with ground state wavefunction
\begin{equation}
	\ket{\Psi_\text{SPT}} =  U_\text{1D} \ket{\Psi_\text{Trivial}} = \smashoperator{\sum_{\st{\sigma^z},\st{\tau^z}}} e^{i\theta_\text{1D}(\sigma^z,\tau^z)} \ket{ \sigma^z,\tau^z},
	\label{eq:SPT_1d_wf}
\end{equation}
with  $e^{i\theta_\text{1D}(\sigma^z,\tau^z)}=\prod (-1)^{\delta_{\downarrow\downarrow}}$. The fact that $H_\text{Trivial}$ and $H_\text{SPT}$ lie in different SPT phases means that transforming one continuously into the other must either break the $\Z_2 \times \Z_2$ symmetry, or close the gap. Both Hamiltonians are short-range entangled, gapped paramagnets and have the same spectrum with periodic boundary conditions. However, with open boundary conditions, they differ at the edge: $H_\text{SPT}$ has spin-$1/2$ gapless edge excitations. We emphasize that the edge modes are topologically protected: they remain when arbitrary perturbations are added to \eqref{eq:SPT_1d} --- so long as the $\Z_2 \times \Z_2$ symmetry is preserved.

\subsection{Gapless $\Z_2 \times \Z_2$ SPT}

Starting from the trivial paramagnet of Eq. \eqref{eq:PM_1d}, one can drive the system to criticality by adding a ferromagnetic interaction for the $\sigma$ spins.
This can also be interpreted as driving the domain walls of $G_{\sigma}$ to criticality.
Explicitly, 
\begin{equation}
	H_\text{gTrivial} = H_\text{Trivial} - \sum_i \sigma_i^z \sigma_{i-1}^z. 
	\label{eq:gTrivial_1d}
\end{equation}
This is a critical Ising chain for $\sigma$ and a trivial paramagnet for $\tau$.
At low energy, one can ignore the gapped $\tau$ degrees of freedom and the criticality is in the Ising universality class.

Using the same local unitary $U_\text{1D}$ as above, we define a gapless SPT system as
\begin{equation}
	H_\text{gSPT} = U_\text{1D} H_\text{gTrivial}  U_\text{1D} = H_\text{SPT} - \sum_i \sigma_i^z \sigma_{i-1}^z. 
	\label{eq:gSPT_1d}
\end{equation}
We will show that, just as with $H_\text{Trivial}$ and $H_\text{SPT}$, $H_\text{gTrivial}$ and $H_\text{gSPT}$ have the same bulk properties but differ at the edge. Namely, $H_\text{gSPT}$ supports topological edge modes. This difference can also be interpreted as a difference of (conformally invariant) boundary condition for the Ising conformal field theory (CFT): the edge modes of $H_\text{gSPT}$ effectively lead to fixed boundary conditions (whereby the spins at the edge are held fixed, either up or down), while $H_\text{gTrivial}$ has a free boundary condition.
Note that fixed boundary conditions for an Ising CFT normally require the symmetry to be explicitly broken at the edge.
Obtaining such boundary conditions for an Ising-symmetric Hamiltonian is therefore highly unusual and a signature of the anomalous character of the boundary properties of $H_\text{gSPT}$.

%With periodic boundary conditions, the spectrum of both $H_\text{gTrivial}$ and $H_\text{gSPT}$ is that of the critical Ising model up to the energy of the gap on the $\tau$ degrees of freedom. However, with open boundary conditions, the edge qbits cause the $H_\text{gSPT}$ spectrum to double with an splitting between the copies that is exponentially small in system size--- see Figure \ref{fig:robustness} (a) for a cartoon. The observable consequence of this doubling is an anomalous magnetization in the $H_\text{gSPT}$ case only.

To see how this comes about, consider the exactly solvable case of $H_\text{gSPT}$ on a semi-infinite chain $i \ge 0$ starting with $\sigma_0$. (For both $H_\text{SPT}$ and $H_\text{gSPT}$, the term $\sigma_0^x\tau_{1/2}$ is disallowed by symmetry, so we start with $\sigma_0^z\sigma_1^z + \sigma_0^z\tau_{1/2}^x\sigma_1^z$.) This Hamiltonian has two exactly degenerate ground states indexed by the edge mode $\sigma^z_0=\pm 1$, denoted $\ket{\Psi_{\text{gSPT}}}_{\pm}$. One easily finds that
\begin{equation}
	\ket{\Psi_{\text{gSPT}}}_{\pm} =  U_\text{1D} \Bigg( \ket{\text{Ising}}_{\pm} \otimes \sum_{\st{\tau^z}} \ket{\tau^z} \Bigg),
	\label{eq:1DgSPTwfct}
\end{equation}
where $U_\text{1D}$ is the unitary defined above restricted to $i>0$ and where $\ket{\text{Ising}}_{\pm}$ are the critical Ising ground states for the $\sigma$ degrees of freedom with fixed boundary spin $\sigma^z_0=\pm \frac{1}{2}$.
Since $U_\text{1D}$ commutes with $\sigma^z_i$, the magnetization $m_i \equiv \left\langle \sigma^z_i \right\rangle$ can be computed for the state $\ket{\text{Ising}}_{\pm}$, for which it is known to decay as $x^{-1/8}$, where $x\propto i$ is the distance from the edge \cite{doi:10.1142/S0217732394002082}.
Of course, the wavefunctions~\eqref{eq:1DgSPTwfct} break the $G_\sigma=\Z_2$ symmetry at the boundary, and the true groundstates will be symmetry-preserving cat states
$
\ket{\Psi_{\text{gSPT}}}_{+} 
\pm
\ket{\Psi_{\text{gSPT}}}_{-}$.
However, as in regular symmetry-breaking, a minute boundary field $h_B \sigma^z_0$ (or bulk field $h \sum_i \sigma_i^z$) is enough to pick either $\ket{\Psi_{\text{gSPT}}}_{+}$ or $\ket{\Psi_{\text{gSPT}}}_{-}$, thereby leading to a non-zero magnetization which decays into the bulk as $x^{-1/8}$.

This is in stark contrast to the gTrivial case where the boundary condition is free, the ground state is non-degenerate and the magnetization is zero, both at the edge and in the bulk \footnote{As always when discussing symmetry-breaking, one should specify how the limits $L\rightarrow \infty$ and $h_B \rightarrow 0$ (or $h \rightarrow 0$) are taken. Strictly speaking, if the boundary field $h_B$ decays slower than $1/\sqrt{L}$, an edge magnetization can occur in the thermodynamic limit due to the fact that a boundary field is a relevant boundary perturbation \cite{doi:10.1142/S0217732394002082}. This is mainly irrelevant in practice to distinguish gSPT and gTrivial since even a field that decays exponentially with $L$ is enough to produce a magnetization in the gSPT case.}.
Note that the bulk magnetization $m_\text{bulk}=\frac1L \sum_i m_i$ also vanishes for gSPT in the limit $L \rightarrow \infty$, although very slowly: $m_\text{bulk} \sim L^{-1/8}$ ($L$ is the system size).

\begin{figure}
	\includegraphics[width=1.0\linewidth]{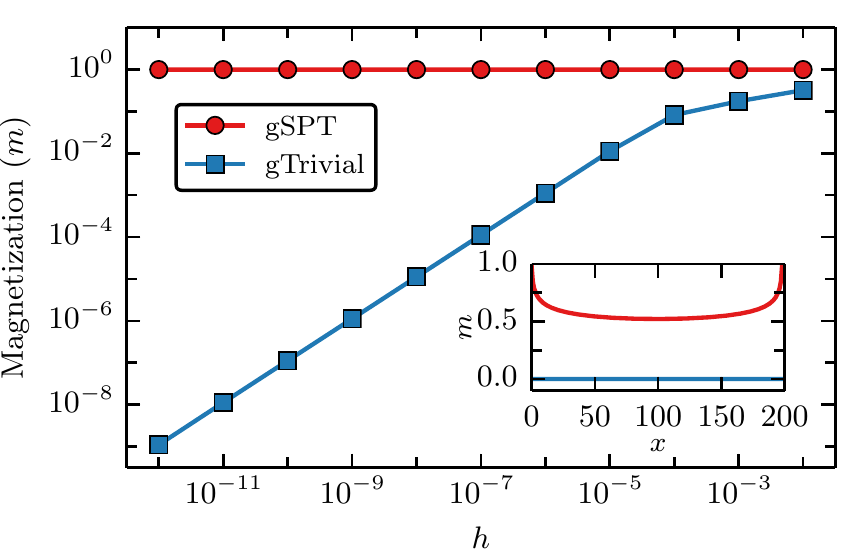}
	\caption{Edge magnetization of the critical $\sigma$ spins for typical gSPT and gTrivial groundstates as a function of a small magnetic field. The groundstates were computed on $L=200$ $\sigma$ spins (and 200 $\tau$ spins) using DMRG, including small but arbitrary symmetry-preserving boundary perturbations. Inset: spatial magnetization profiles for a field $h=10^{-10}$.  }
	\label{fig:1d}
\end{figure}

Using standard density matrix renormalization group techniques~\cite{white92,schollwoeck}, we numerically compare the typical magnetization profile for gSPT and gTrivial systems with open boundary conditions. We include small but arbitrary symmetry-preserving boundary perturbations, and a small $ g_\tau  \sum_i \tau_{i-1/2}^z \tau_{i+1/2}^z$ term that gives a non-zero correlation length to the gapped $\tau$ spins. In the presence of a magnetic field much smaller than the CFT finite size gap, we find a clear qualitative difference between gSPT and gTrivial systems (Fig.~\ref{fig:1d}).

The properties of $H_\text{gSPT}$ are robust and not a product of fine-tuning. They are stable in the presence of any symmetry-preserving perturbations, as long as the $\tau$ gap is not closed and the $\sigma$ spins remain critical. The entire phase boundary between the non-trivial SPT (paramagnet) to a ferromagnet has the character of a gSPT, and we expect our conclusions to broadly apply to more general phase transitions between SPT and broken-symmetry phases. 

\begin{figure}
	\includegraphics{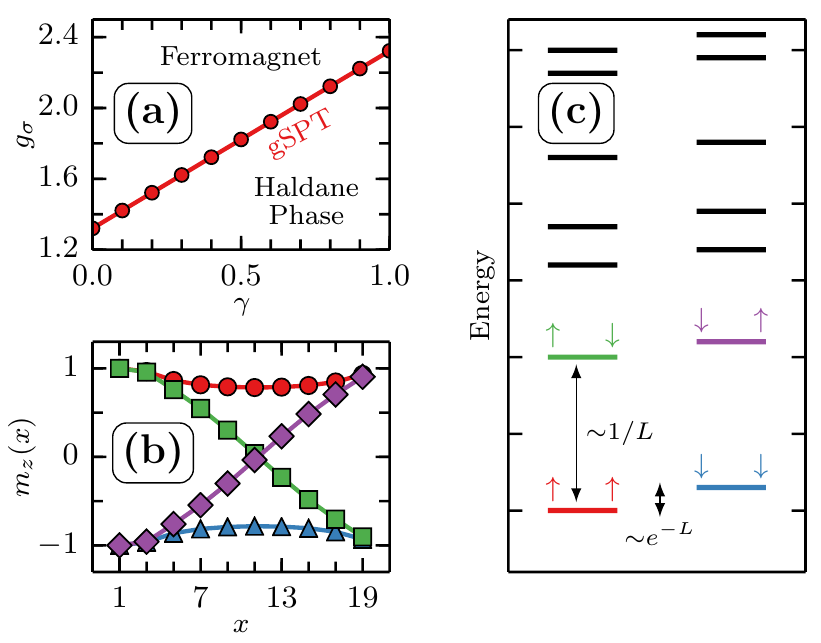}
	\caption{(a) Phase diagram showing a gSPT line separating the Haldane and ferromagnetic  phases obtained from ED on 12 and 16 sites with $\Delta_\tau = 10$, $g_\tau = u_\sigma = 0.1$, $u_\sigma = \delta = 0.2$.  (b) Magnetization profiles of the lowest four eigenstates via ED on 20 sites with the same parameters as (a), but fixing $\gamma = 0.1$, which implies $g_\sigma = g_\sigma^c \approx 1.421$ at the gSPT point. To break the symmetry, a small magnetic field $\sim{\rm e}^{-L}$ in the $z$ direction is applied. (c) Cartoon spectrum of $H_\text{gSPT}'$. Colors of states correspond to magnetization profiles.} 
		\label{fig:robustness}
\end{figure}

We add several types of perturbations to $H_\text{gSPT}$ and consider the generalized Hamiltonian 
\begin{equation}
	H_\text{gSPT}' = U_\text{1D} H_\text{gTrivial}' U_\text{1D} + \delta(\sigma_0^x + \tau_{\frac{1}{2}}^x + \sigma_{L-1}^x + \tau_{L- \frac{1}{2}}^{x} ),
\end{equation}
where 
\begin{align*}
	& H_\text{gTrivial}' \ =\ - \sum_i \sigma_i^x + g_\sigma \sigma_i^z \sigma_{i+1}^z + u_\sigma \sigma_i^x \sigma_{i+1}^x\\
  &\ -  \sum_i \Delta_\tau\left(\tau_{i-\frac{1}{2}}^x + g_\tau \tau_{i-\frac{1}{2}}^z \tau_{i+\frac{1}{2}}^z + u_\tau \tau_{i-\frac{1}{2}}^x \tau_{i+\frac{1}{2}}^x\right) + \gamma \sigma_{i}^x\tau_{i+\frac{1}{2}}^x.
\end{align*}
Here, $\delta$ parametrizes additional terms at the edges, $g_\tau$ gives a non-zero correlation length for the $\tau$ spins, $u_\sigma$ and $u_\tau$ are interaction terms for the $\sigma$ and $\tau$ spins to take them away from integrable points, and $\gamma$ couples the $\sigma$ and $\tau$ sectors. 
The parameter $\Delta_\tau$ controls the gap of the $\tau$ spins, which is used to improve finite-size convergence in exact diagonalization (ED). We choose the parameters $u_\sigma$,  $u_\tau$, $g_\tau$, and $\gamma$ so that the $\tau$ spins remain gapped, deep in their paramagnetic phase, and we tune a single parameter $g_\sigma$  to bring the $\sigma$ spins to criticality. 
%Integrating out the gapped $\tau$ spins, we get a mean-field approximation that the critical value $g^c_\sigma$ of $g_\sigma$ should scale as $g^c_\sigma = g^c_\sigma(\gamma=0) + \gamma \langle \tau^x \rangle$ as a function of the coupling $\gamma$ between the $\tau$ and $\sigma$ spins. 
Using exact diagonalization, we identified the location of the new critical point by studying the finite size crossing of the gap of the system (see phase diagram in Fig.~\ref{fig:robustness}(a)).  We have verified that $H_\text{gSPT}'$ has gapless edge modes and anomalous magnetization for the parameter ranges $0 \le u_\sigma, g_\tau, u_\tau, \delta \le 0.2$ and $0 \le \gamma \le 1$. 

Away from the exactly solvable limit described in the previous section, the exact degeneracy of the groundstate is lifted by quantum fluctuations. There are two nearly degenerate groundstates which correspond to cat state superpositions $\ket{+_1+_L} \pm \ket{-_1-_L}$  of the edges modes. The splitting between these two cat states with lowest energy remains generically protected by the gap of the $\tau$ spins and is exponentially small in system size, well below the finite size CFT gap that scales as ${\sim}1/L$. The first excited states are also cat states corresponding to the configurations $\ket{+_1-_L}$, $\ket{-_1+_L}$ of the edge modes. They are power-law split from the two groundstates because the anti-aligned edge modes induce a change of boundary condition (Fig.~\ref{fig:robustness}(c)). In the CFT language, this corresponds to the insertion of a boundary condition changing operator~\cite{cardy_conformal_1984,cardy_boundary_1989}, which leads to a finite-size gap $\frac{\pi v_F}{2L}$ for a system of size $L$~\cite{0305-4470-17-7-003} with $v_F$ the Fermi velocity. Figure~\ref{fig:robustness}(b) shows the anomalous magnetization of the low-lying eigenstates for non-trivial parameter values, consistent with the above picture.

In conclusion, this system provides an example of a 1+1D gapless SPT as a decorated critical Ising model. We showed that   the anomalous edge properties of $H_\text{gSPT}$ are robust, and do not require any additional fine-tuning beyond making the $\sigma$ spins critical.
This gSPT state can also be interpreted as a quantum critical point between a non-trivial SPT and a ferromagnet, although we emphasize again that our general construction also applies to gapless phases, including Luttinger liquids in 1D (see below). 
The presence of exotic edge properties at this transition stands in contrast to previous works on transitions between trivial and non-trivial SPTs~\cite{2015arXiv151107460T,Tsui2015330,Lokman3,PhysRevB.93.195141}. This should admit straightforward generalizations to Potts models and parafermions in the case of a $\Z_N \times \Z_N$ symmetry. 

Numerically, the anomalous edge magnetization even appears to survive disorder. Because of the unitary twist relating $H_\text{gTrivial}$ and $H_\text{gSPT}$, the stability of the gSPT critical point against disorder is determined by the Harris criterion for the gTrivial system (disorder is irrelevant if the correlation length exponent satisfies $\nu \geq 2/d$), and by the gap of the $\tau$ spins. gSPTs should therefore be as stable against disorder as their gTrivial counterparts before applying the unitary twist. Moreover, even if disorder is relevant, we expect that disordered examples of gSPT systems could be uncovered by studying the boundary physics of twisted infinite randomness critical points~\cite{PhysRevLett.69.534}. This could lead to ``topological'' random singlet phases both at zero temperature~\cite{PhysRevLett.69.534,PhysRevB.50.3799} and in the context of many-body localization~\cite{PhysRevX.4.011052,PhysRevLett.112.217204,PhysRevLett.114.217201}. Furthermore, the possible presence of a strong zero mode \cite{1742-5468-2012-11-P11020,PhysRevB.90.165106,1751-8121-49-30-30LT01} in such models should be investigated.

\begin{figure*}
	\includegraphics[scale=1.0]{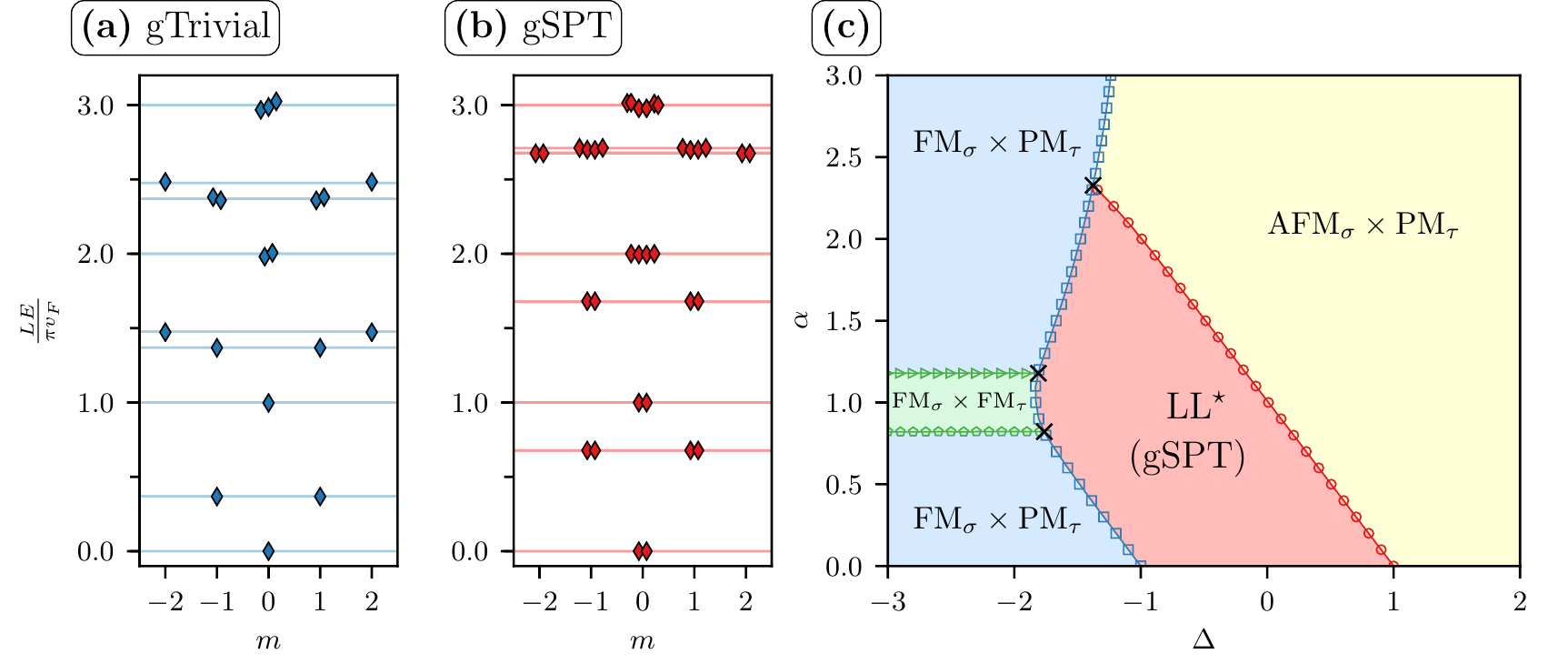}
	\caption{(a) Spectrum of $H_{\text{gTrivial}}^{\text{LL}}$. (b) Spectrum of the gSPT $H_{\text{gSPT}}^{\text{LL*}}$. For both cases, and spectra are normalized to be able to read off CFT operator dimensions. The conformal blocks are labelled by the magnetic charge sector $m$ and spaced horizontally, and small horizontal spacings show degenerate eigenvalues (up to exponential splitting). One can see that the states in the gSPT case are all doubly degenerate, due to the edge modes, but also that operator dimensions have changed relative to the gTrivial case. The numerical spectra were computed via DMRG \cite{ITensor} on up to 32 sites with finite-size scaling, and the solid lines correspond to the exponents expected from boundary CFT using $\Delta_{\rm eff} = - \cos \pi g$~\cite{toAppear}. To improve convergence, the gap on the paramagnetic sector was  increased from one to ten. (c) The phase diagram of $H_{\text{gSPT}}^{\text{LL*}}$, as computed via DMRG \cite{ITensor}. Each line denotes a different eigenvalue crossing which accompanies a phase transition, and black crosses denote multicritical points. The Hamiltonian parameters used are $\Delta = -0.5, \alpha = 0.1$, $g_\tau = 0.3$, $u_\tau = 0.1$ for (a), (b) and (c).
}
	\label{fig:xxz_star}
\end{figure*}

\subsection{$U(1) \rtimes \Z_2 \times \Z_2$ gSPT phase in 1D}
\label{secLLstar}

The presence of a quantum critical point in the preceding example is a special case; our construction can be applied not only to gapless points, but equally well to lines or phases. To emphasize the generality of our construction, we now present a gapless SPT phase in 1D. 
%	\sout{
%		So far, our example of gSPT system~\eqref{eq:gSPT_1d} was restricted to a quantum critical point, and the presence of anomalous boundary physics can be interpreted as a general feature of quantum phase transitions separating SPT and broken-symmetry phases. To emphasize the generality of our construction which is by no means restricted to critical points, we discuss here how to construct a gSPT Luttinger liquid phase with anomalous edge modes in 1D. }}

%		\Dan{\sout{In order to construct an example of gapless SPT phase,}} 
		We start from a (``gTrivial'') gapless phase in 1D -- a Luttinger liquid~\cite{giamarchi2003quantum}. The systematic nature of our construction allows us to closely follow the $\Z_2 \times \Z_2$ example above, but we will enforce an additional $U(1)$ symmetry on the gapless $\sigma$ spins in order to lock them into a Luttinger liquid phase. We start from the Hamiltonian
\begin{equation}
	\begin{aligned}
	& H^{\rm LL}_\text{gTrivial} \ =\ \sum_i \sigma_i^x \sigma_{i+1}^x + \sigma_i^z \sigma_{i+1}^z + \Delta \sigma_i^y \sigma_{i+1}^y \\
	 &- \sum_i \tau_{i-\frac{1}{2}}^x + g_\tau \tau_{i-\frac{1}{2}}^z \tau_{i+\frac{1}{2}}^z + u_\tau \tau_{i-\frac{1}{2}}^x \tau_{i+\frac{1}{2}}^x + \alpha  \sigma_i^y \tau_{i+ \frac{1}{2}}^x \sigma_{i+1}^y,\\ 
\end{aligned}
	\label{eq:H_gTrivialXXZ}
\end{equation}
which describes gapless $\sigma$ spins (XXZ model) coupled through the $\alpha$ term to gapped Ising $\tau$ spins deep in their paramagnetic phase ($u_\tau$ and $g_\tau$ are small). (Note that in contrast to the usual convention for the XXZ spin chain, $\Delta$ adjusts the magnitude of the $\sigma^y \sigma^y$ interaction, to make the symmetries more convenient.) This has a $U(1) \rtimes \Z_2^{(\sigma)} \times \Z_2^{(\tau)}$ global symmetry, generated by $U_\theta = \prod_i e^{i \theta \sigma^y_i}$, $\mathcal{C}_\sigma = \prod_i \sigma_i^x$ and $\mathcal{C}_\tau = \prod_i \tau_{i+1/2}^x$ respectively. 
%\Dan{The interaction $\alpha$ is chosen as the simplest coupling between the $\sigma$'s and $\tau$'s compatible with the symmetries.} 
Assuming that $\alpha$ is small, the gapped $\tau$ spins can be integrated out to renormalize the anisotropy parameter $\Delta_{\rm eff} = \Delta - \alpha \langle \tau^x \rangle$. The resulting $\sigma$ spins are gapless for $-1<\Delta_{\rm eff} \leq 1$, and form at low energy a (single channel) Luttinger liquid phase with effective Lagrangian 
\begin{equation}
{\cal L} = \frac{g}{4 \pi} (\partial_\mu \phi )^2,
\end{equation}
with $\Delta_{\rm eff} = - \cos \pi g$, $\phi$ a compact boson with unit compactification radius. (We set the Fermi velocity to $v_F=1$ for simplicity.)

Upon applying the unitary twist~\eqref{eqUnitaryTwist1D} and following the same steps as above, one can readily show that the twisted Hamiltonian $H^{\rm LL^\star}_\text{gSPT} = U_\text{1D} H^{\rm LL}_\text{gTrivial} U_\text{1D} $ has edge modes in the limit $\alpha = g_\tau = u_\tau=0$ \footnote{Note that the $U(1)$ symmetry also becomes twisted in the process.}. These topological edge modes are robust and persist away from this special limit as long as the gap of the $\tau$ spins does not close. Similarly to the $\Z_2 \times \Z_2$ example above, the edge modes can be thought of as inducing a spontaneous edge magnetization along the $z$ direction, which in turn induces a change of (conformally invariant) boundary conditions~\cite{0305-4470-31-12-003}. Using standard bosonization techniques, the edge modes can be seen to lead to a doubly-degenerate spectrum of boundary critical exponents that can be obtained from the gTrivial case through the substitution $g \to \frac{1}{4 g}$~\cite{toAppear}. (Note that this is in sharp contrast with the $\Z_2 \times \Z_2$ gSPT discussed above where the edge modes only led to degeneracies and did not modify the value of the critical exponents). We have checked these predictions and the robustness of this topological Luttinger liquid (gSPT) phase using exact diagonalization and DMRG calculations~\cite{toAppear}.

We emphasize that contrary to other examples of topological Luttinger liquids previously discussed in the literature~\cite{PhysRevB.91.235309,2017arXiv170402997J}, our construction does not rely on the spin charge separation property of Luttinger liquids. Instead, our decorated domain wall construction provides us with a systematic way of generating strongly-interacting gapless SPT phases, while making their topological transparent in clear analogy with gapped SPT systems.

\section{Two dimensions}

To showcase the range of our general construction, our second example is a more involved system in 2D. However, the construction is parallel to the last section. We first define the model and then proceed to analyze its behavior in subsequent sections.

This example has a $G_\sigma \times G_\tau =\Z_2 \times (\Z_2 \times \Z_2)$ symmetry where the domain walls of $G_{\sigma}=\Z_2$ will be decorated with (gapped) one-dimensional SPT states protected by $G_{\tau}=\Z_2 \times \Z_2$.
Let $A$ be a lattice whose sites host $\sigma$ spins, with symmetry ${\cal C}_A = \prod_{a \in A} \sigma_a^x$.
The $\tau$ spins live on the sites of the dual (face-centered) lattice of $A$, called $A^\star$, which we assume bipartite so that $A^\star = B \cup C$ with symmetries given by ${\cal C}_B = \prod_{b \in B} \tau_b^x$ and ${\cal C}_C = \prod_{c \in C} \tau_c^x$.
We will further assume a symmetry exchanging $B$ and $C$. This can be realized either on triangular, or Union Jack lattices, as shown in Fig.~\ref{fig:lattices}. A ``trivial'' paramagnetic state can be obtained as an equal-weight superposition of all classical configurations of spins, with parent Hamiltonian $H_\text{Trivial} = -\sum_{a \in A} \sigma_a^x-\sum_{a^\star \in A^\star} \tau_{a^\star}^x$ and ground state wavefunction
\begin{equation}
	\ket{\Psi_\text{Trivial}} = \sum_{\st{\sigma^z}, \st{\tau^z}} \ket{\sigma^z, \tau^z}.
	\label{eq:trivial_GS}
\end{equation}

Following the well-known construction \cite{Chen2011,PhysRevB.93.155131}, a parent Hamilonian for a non-trivial $\Z_2^3$ SPT is given by $H_\text{SPT} = U_\text{2D} H_\text{Trivial} U_\text{2D}$, where $U_\text{2D} = \prod_{\Delta_{ijk}} \rm{CCZ}_{ijk}$ is a local unitary operator that applies a three-qubit operator on each triangle of three neighboring ABC sites (see  Fig.~\ref{fig:lattices}).
This control-control-$Z$ operator gives a $-1$ for three down spins and $+1$ otherwise: $\rm{CCZ}_{ijk} = (-1)^{\delta_{\downarrow\downarrow\downarrow}}$.
%\begin{equation}
%\rm{CZ}_{ijk} \ket{\sigma^z_i \tau^z_j \tau^z_k} = (-1)^{\delta_{\downarrow\downarrow\downarrow}} \ket{\sigma^z_i \tau^z_j \tau^z_k}.
%\end{equation}
One can check that, for each edge $\left\langle jk \right\rangle$ of $A^{\star}$ that hosts a $\sigma$ domain wall, $U_{2D}$ applies the 2-qubit control-$Z$ operator $\rm{CZ}_{jk}$ on the $\tau$ spins $j$ and $k$.
This unitary therefore applies $U_\text{1D}$ to the $\tau$ spins living on each domain wall of the $\sigma$ spins, thereby decorating them with a 1D SPT chain protected by a $ \Z_2 \times \Z_2 $ symmetry.
%
%on the two sites of that edge
%
%
%$U_\text{2D}$ decorates each edge of the dual lattice
%
%One can check this decorates each domain wall of $A$ with a 1D SPT chain protected by a $ \Z_2 \times \Z_2 $ symmetry.
%On the triangular and squares lattices, the product is over all triangles with one $A$ spin and two adjacent $A^*$ spins, and $\theta = \pi$ if all three spins point down, and $\theta=0$ otherwise. 
%Then $H_\text{SPT} = U H_\text{PM} U$ is the parent Hamiltonian of a non-trivial $\Z_2^3$ SPT. 
Explicitly, we have
\begin{equation}
	\ket{\Psi_\text{SPT}} = U_\text{2D} \ket{\Psi_\text{Trivial}} = \smashoperator{\sum_{\st{\sigma^z}, \st{\tau^z}}} e^{i\theta_\text{2D}(\sigma^z, \tau^z)}  \ket{\sigma^z, \tau^z},
	\label{eq:SPT_GS}
\end{equation}
where the $\sigma$ and $\tau$ spins are now coupled through the phase factor $e^{i\theta_\text{2D}(\sigma^z, \tau^z)}$ which takes care of the domain wall decoration:

\begin{equation}
	e^{i\theta_\text{2D}} = \prod_{\st{\text{dw}}} e^{i\theta_\text{1D}(\tau_\text{dw})}.
	\label{eq:2DSPTPhase}
\end{equation}
where the product is over the domain walls of $\sigma^z$, denoted by $\st{\text{dw}}$, and where $e^{i\theta_\text{1D}(\tau_\text{dw})}$ is defined in the previous section, and applied to the $\tau$ spins living on a given domain wall $\text{dw}$.

% defined by $ U_\text{2D}   \ket{\sigma^z, \tau^z} = e^{i\theta(\sigma^z, \tau^z)}  \ket{\sigma^z, \tau^z} $.

%This is a garden-variety --- and hence gapped --- SPT described by the classification of 2+1d SPTs and has been studied by several authors \cite{Chen2011,PhysRevB.93.115105,PhysRevB.93.155131}.
For a region with an edge of $B$ and $C$ sites, $U_\text{2D}$ does not modify the $G_{\tau}$ symmetry generators ${\cal C}_B$ and ${\cal C}_C$, but does lead to additional boundary terms in $U_\text{2D}{\cal C}_A U_\text{2D} = \prod_{a \in A} \sigma_a^x \prod_{\partial A^* } {\rm CZ}$ where $\rm{CZ}$ is a control-$Z$ gate giving a (-1) factor if two successive $B$ and $C$ boundary spins are down. We can write down the edge theory of this $\Z_2^3$ SPT following Levin and Gu~\cite{Levin2012} by including all terms allowed by the symmetries, such as $\tau^z_{B,i-1} \tau^x_{C,i} \tau^z_{B,i+1} +  \tau^x_{C,i} $, $\tau^z_{B,i-1} \tau^z_{B,i+1} $ and $B \leftrightarrow C$ permutations. 

Using standard duality arguments, the edge theory can be thought of as two coupled Ising models tuned to their self-dual critical points (also known as the Ashkin-Teller model~\cite{PhysRev.64.178}). After bosonization, the edge excitations can be described by a Luttinger liquid~\cite{giamarchi2003quantum} with central charge $c=1$ at the electromagnetic self-dual point
\begin{equation}
	{\cal L}_\text{SPT}^{\rm edge} =   \frac{1}{4\pi} \left( \partial_\mu \phi \right)^2  - \lambda \left( \cos 2\phi + \cos 2 \theta \right),
	\label{eq:SPT_edge}
\end{equation}
where $\phi,\theta$ are compact conjugate bosonic fields with unit compactification radius. The edge is protected by the symmetries  $\phi \to \pm \phi + \pi$, $\theta \to \pm \theta + \pi$, and $\phi \leftrightarrow \theta$ (note that the last symmetry is generated by the symmetry exchanging B and C spins). The vertex operators $ \cos 2\phi$ and  $\cos 2\theta$ correspond to products of the energy operators of the two Ising models. They are marginal perturbations that can be absorbed by renormalizing the Luttinger parameter and the sound velocity \cite{Lecheminant2002502}. 

\subsection{gSPT wavefunction}

Now we tune the $\sigma$ spins to criticality by imposing the constraint that the domain walls of the $\sigma^z$-spins must be fully-packed loops (FPL) \cite{PhysRevLett.72.1372,0305-4470-29-20-007}. 
On the triangular lattice, this corresponds to a natural physical constraint: the allowed $\sigma^z$ states are the maximally anti-ferromagnetic ones which, due to frustration, are known to have extensive degeneracy and power law correlations \cite{PhysRev.79.357}.
On the square lattice, the FPL constraint is equivalent to the ice rule of the 6-vertex model~\cite{0305-4470-29-16-001}. 
For concreteness, we focus on the triangular lattice, for which the fully-packed loops live on the dual honeycomb lattice. 
For a given site $a \in A$, let $P_a$ be the projector onto allowed configurations (i.e. configurations that respect the constraint for the six triangles surrounding $a$) and let $P'_a$ be the projector onto allowed configurations for which $a$ is ``resonant'' (i.e. configurations that would still respect the constraint after flipping $\sigma^z_a$).
Since $P_a$ and $P'_a$ are only functions of the $\sigma^z$ operators on $a$ and its neighbors (on the $A$ lattice), they are local operators.
Then the gTrivial Hamiltonian is (still with $G_{\sigma} \times G_{\tau}$ symmetry)
\begin{equation}
	H_\text{gTrivial} = \sum_{a \in A} \Lambda (1 - P_a) + \sum_{a \in A} P'_a \left( 1 - \sigma_a^x \right) P'_a - \sum_{a^\star \in A^\star} \tau_{a^\star}^x,
	\label{eq:gTrivial_2d}
\end{equation}
where $\Lambda \rightarrow \infty$ is an energy cost to penalize configurations that do not respect the constraint. To find the exact ground state, note that the $\tau$ spins are completely decoupled from the $\sigma$ spins. For the $\sigma$ degrees of freedom, we can follow the standard argument due to Rokhsar and Kivelson~\cite{Rokhsar:1988zz}. The $\sigma$ part of $H_\text{gTrivial}$ is a sum of projectors and is therefore positive semi-definite. 
%This Hamiltonian is, up to a constant, a sum of projectors and is therefore positive semi-definite.
%Because $H_\text{gTrivial}^2 = 2H_\text{gTrivial}$, this is positive semi-definite. 
%Since the $\sigma$ and $\tau$ spins are decoupled, we can follow the standard argument due to Rokhsar and Kivelson~\cite{Rokhsar:1988zz} to treat the $\sigma$ spins.
%An equal-weight superposition over all $\sigma^z$ states that satisfy the constraint (denoted $\overline{\st{\sigma^z}}$) has zero energy.
Thus, the (unnormalized) state
\begin{equation}
	\ket{\Psi_\text{gTrivial}} = \sum_{\overline{\st{\sigma^z}},\st{\tau^z}} \ket{\overline{\sigma^z},\tau^z},
	\label{eq:g_trivial_GS}
\end{equation}
an equal-weight superposition over all $\sigma^z$ states that satisfy the constraint (denoted $\overline{\st{\sigma^z}}$) times a paramagnetic state for the $\tau$ spins, has zero energy under the $\sigma$ part of $H_\text{gTrivial}$ and is hence an exact ground state.
%an equal-weight superposition over all allowed $\sigma^z$ states (noted $\overline{\sigma^z}$) tensor product a trivial paramagnet for the $\tau$ spins, has zero energy and is hence an exact ground state. 
Equal-time $\sigma^z$-correlation functions in the ground state are described by correlation functions in the 2D FPL model with loop fugacity $n=1$ \cite{PhysRevLett.72.1372,0305-4470-29-20-007}, or equivalently by correlation functions in the zero temperature triangular lattice Ising antiferromagnet \cite{PhysRev.79.357}.

Using standard mappings onto dimers and height models~\cite{PhysRevLett.72.1372,0305-4470-29-20-007}, the continuum limit of the 2D FPL model can be identified as a $c=1$ compact boson CFT ${\cal L} = \frac{g}{4 \pi} (\nabla \varphi)^2 - \gamma \cos 3 \varphi$ with $g = \frac{1}{2}$ and $\varphi \equiv \varphi + 2\pi$ so the perturbation $ \cos 3 \varphi$ has scaling dimension $\Delta=9$ and is irrelevant. Following Refs. \cite{Ardonne2004493,PhysRevB.69.224415,fradkin2013field}, we quantize this theory to identify the $2+1d$ effective field theory describing the low energy physics of $H_\text{gTrivial}$ as the $z=2$ quantum Lifshitz model (QLM) with (Euclidian) Lagrangian density 
\begin{equation}
	{\cal L}_{\rm QLM}^{\rm bulk}= \frac{1}{2}\left[ \left( \partial_\tau \varphi \right)^2 + k \left(  \nabla \varphi \right)^2 + \kappa^2 \left(  \nabla^2 \varphi \right)^2 \right] - \gamma \cos 3 \varphi ,
	\label{eq:QLM}
\end{equation}
tuned to $k=0$ with $\kappa = 1/(8\pi)$ to reproduce the equal-time antiferromagnetic spin correlations on the triangular lattice. This constitutes an effective field theory for the gTrivial order on a closed manifold, and is manifestly gapless. Equivalently, one can also think of this quantum critical point in terms of a dual $U(1)$ gauge theory with a quadratic photon mode \cite{fradkin2013field}.

The stability of this quantum critical point has been studied in various contexts~\cite{PhysRevB.69.224415,PhysRevB.69.224416,PhysRevB.87.085102} and depends on crystalline symmetries, with the important relevant perturbations in our case being magnetic operators breaking the FPL constraint, and $ k \left(  \nabla \varphi \right)^2$ that makes $\cos 3 \varphi$ relevant and opens up a gap (this can be equivalently interpreted as the instability of the deconfined phase of $U(1)$ gauge theories in $2+1d$~\cite{PhysRevD.19.3682}.) In the following, we will assume the bulk is tuned to this quantum Lifshitz critical point. The  $\Z_2^3$ symmetry discussed above acts trivially on $\varphi$ but the theory~\eqref{eq:QLM} has additional crystalline symmetries corresponding to three-fold rotations $\varphi \to \varphi + 2 \pi/3$ and inversion $\varphi \to -\varphi$. A similar field theory can be obtained on the square lattice~\footnote{For fully-packed loops on the square lattice, one can use a mapping onto a 6-vertex model with anisotropy parameter $\Delta=-\frac{1}{2}$ which can be described by a free boson CFT with parameter $g=\frac{1}{3}$. This theory has an additional symmetry  $\varphi \to \varphi + \pi$ due to the bipartite nature of $A$ that corresponds to flipping every other spin on the $A$ lattice. }.

% has a additional symmetry $\varphi \to \varphi + \pi$ that corresponds to flipping every other spin on the $A$ lattice: this additional sublattice symmetry comes from the bipartite nature of $A$.

An example of 2+1d gapless SPT order is now obtained by decorating $H_\text{gTrivial}$,
	\begin{equation}
H_{\text{gSPT}} = U_\text{2D} H_\text{gTrivial} U_\text{2D}.
\label{eq:gSPT_2d}
\end{equation}
 Its ground state is simply
\begin{equation}
	\ket{\Psi_\text{gSPT}} = U_\text{2D} \ket{\Psi_g} =  \smashoperator{\sum_{\overline{\st{\sigma^z}},\st{\tau^z}} } e^{i\theta_\text{2D}(\overline{\sigma^z}, \tau^z)}  \ket{\overline{\sigma^z},\tau^z}.
	\label{eq:g_SPT_GS}
\end{equation}
%One can check that $\Z_2 \times G$ is a symmetry of both ground states. 

We will now argue --- crucially --- that the critical wavefunction $\ket{\Psi_\text{gSPT}}$ has an extra gapless edge mode compared to $\ket{\Psi_\text{gTrivial}}$, and that this edge mode is protected. This behavior is a hallmark of SPT order, and must be treated with care in this gapless context. We will therefore present three independent arguments for it: (1) effective field theory and boundary renormalization group (RG), (2) bulk-boundary correspondence, and (3) entanglement spectrum calculations with numerics. Each argument separately confirms a gapless $c=1$ edge in the gTrivial case and a gapless $c=2$ edge in the gSPT case. 

\subsection{Edge Field Theory}
%combine the two field theories into an ABC field theory and look at the boundary
%RG analysis to see what can appear

We first consider the edge modes of the (topologically trivial) gapless state $\ket{\Psi_\text{gTrivial}}$, Eq.~\eqref{eq:g_trivial_GS}, for which Eq.~\eqref{eq:QLM} describes the bulk behavior of the $\sigma$ spins.  Because the  boundary conditions for the spins are free, we consider Neumann boundary conditions for the field $\varphi$. (Note that Dirichlet boundary conditions $\left. \varphi \right|_\partial = 0$ for the QLM are RG unstable and flow to Neumann as the normal derivative boundary perturbation $\left. (\partial_n \varphi)^2 \right|_\partial$ has scaling dimension $\Delta=2 < z+1=3$ and is therefore relevant.) However, it is important that even though the relativistic $z=1$ term in Eq.~\eqref{eq:QLM} is tuned to $k=0$, such quadratic  terms have no reason to be set to zero at the edge without additional fine tuning. At the boundary, one should therefore add a lateral derivative boundary term $V \sim \delta^2 \int_{\partial} d\tau \, dx_\parallel  \left( \partial_{x_\parallel} \varphi \right)^2$  to the action. Here, $x_\parallel $ is the coordinate along the edge and  $\delta$ is a non-universal parameter. The boundary perturbation $ \left( \partial_{x_\parallel} \varphi \right)^2$ is relevant and we conjecture that in the IR, it endows the edge with $z=1$ dynamics (forgetting the slower $z=2$ bulk dynamics). This leads to the effective low energy action for the edge theory of the QLM
\begin{equation}
	S_\text{QLM}^\text{edge} = \frac{1}{2} \int d\tau \, dx_\parallel  \left[ \left( \partial_\tau \varphi \right)^2 + \delta^2 \left( \partial_{x_\parallel} \varphi \right)^2 \right] + \cdots
	\label{eq:QLM_edge}
\end{equation}
where the dots represent less RG-relevant terms.  This is the action of a 1+1d compact boson CFT with central charge $c=1$. We emphasize that the effective Luttinger parameter is non-universal and set by the value of $\delta$, which depending on microscopic parameters could lead to a gapped edge because of the cosine terms dropped in~\eqref{eq:QLM_edge}. The existence of this edge has nothing to do with the SPT and indeed the symmetries ($\mathcal{C}_A, \mathcal{C}_B, \mathcal{C}_C$) act trivially on $\varphi$. The $\sigma$ spins therefore have a $z=2$ bulk with diffusive dynamics and can have a $z=1$ edge with ballistic dynamics. To our knowledge, edge modes for gapless systems have been very rarely discussed in the literature~\cite{PhysRevB.82.085102,PhysRevB.93.174414,PhysRevB.92.165125,PhysRevB.94.075101,2012arXiv1206.1332G}. The presence of an edge is confirmed numerically below.

We now turn to the edge theory of the gSPT wavefunction~\eqref{eq:g_SPT_GS}. Upon integrating out the gapped $\tau$ degrees of freedom, we expect the bulk low-energy theory to be described by~\eqref{eq:QLM} with $k=0$, where the boundary actions~\eqref{eq:SPT_edge} and~\eqref{eq:QLM_edge} (both with non-universal Luttinger parameters) are coupled through all symmetry-allowed perturbations. The essential point is that the SPT $\Z_2^3$ symmetry acts trivially on $\varphi$, so perturbations such as $\cos(\varphi - \phi)$ that could generically gap out the edge are not allowed by symmetry. Intuitively, coupling the two edge theories does not increase the number of symmetry-allowed perturbations, since they are protected by distinct symmetries.
%\Romain{give them more chances of being gapped since they are protected by distinct symmetries, so they are effectively decoupled at low energy}. 
There is therefore a finite range of Luttinger parameters for which the edge is gapless with central charge $c=2$. Moreover, the SPT part of the edge, described by Eq.~\eqref{eq:SPT_edge}, is symmetry protected as it can only be gapped out by condensing $\phi$ or $\theta$, thereby spontaneously breaking the $\Z_2^3$  symmetry. A related mechanism for non-interacting gapless topological superconductors and insulators has been discussed in Ref.~\cite{2012arXiv1206.1332G}.

Investigating to what extent these edge modes leak into the gapless bulk is a complicated task. In analogy with the 1D case, at least the $\tau$-component of the edge should be exponentially localized despite the critical bulk. Even though the domain walls themselves are critical, the 1+1d SPT chains that live on them are still gapped. 
Therefore, when a domain wall ends at a boundary, there is a free spin $1/2$ living at the end point (see Fig.~\ref{fig:cartoon}) whose only way to move towards the bulk is along the domain wall, which is forbidden by the 1+1d SPT gap. 
We thus expect the free edge spins to be exponentially localized, where the localization length is given by the gap on the 1+1d SPT chains.

%Therefore \Dan{for each domain wall that intersects the edge, there is a free spin $1/2$ at the boundary (see Fig.~\ref{fig:cartoon}). The spins' movement is constrained to the domain wall, and forbidden from penetrating far into the bulk by the 1+1d SPT gap.} Therefore, we expect the \Dan{edge modes} to be exponentially localized, where the localization length is given by the gap on the 1+1d SPT chains.

%\Dan{This should be double checked and perhaps fleshed out.}
%We may form an effective field theory for the gapless SPT order by combining \eqref{eq:SPT_edge}, \eqref{eq:QLM}, and \eqref{eq:QLM_edge},  and adding all perturbations allowed by symmetry. However, \eqref{eq:SPT_edge} is only non-trivial at the boundaries, so it is sufficient to combine the boundary field theories. By performing an RG analysis, we find that the combined edge theory consists of two uncoupled Luttinger liquids with renormalized parameters. The only relevant perturbations are magnetic operators, whose inclusion will explicitly break the FPL constraint. Therefore the gapless SPT order has a $c=2$ edge and requires no fine-tuning beyond the FPL constraint present in the bulk.

\subsection{Bulk-boundary correspondence}
%\Thomas{[I replaced strange correlators with bulk-boundary correspondence as I am afraid strange correlators might sound unfamilar to most readers]}

The bulk-boundary correspondence for fractional quantum Hall (FQH) states~\cite{Fubini1991,MooreRead1991,Blok1992615,NayakRevMod2008} is a powerful technique whereby a FQH wavefunction is written as a correlator in a CFT.
When it is unitary, the CFT also gives the edge and entanglement spectra~\cite{PhysRevLett.101.010504,PhysRevB.84.205136,PhysRevB.86.245310}.
This correspondence was recently extended to SPT wavefunctions~\cite{PhysRevB.93.115105} (see also \cite{2016arXiv160606402C}) and we build on this result in this section.

Starting from an SPT wavefunction, a convenient way of identifying its underlying CFT is to compute the ``strange correlator theory'' $Z_{\Psi_\text{SPT}} \equiv \braket{\Psi_\text{Trivial}|\Psi_\text{SPT}}$~\cite{Cenke2014}.
The idea behind this theory is that correlators of the type $\braket{\Psi_\text{Trivial}|\mathcal{O}_1 \mathcal{O}_2 |\Psi_\text{SPT}}$ measure observables on an edge in imaginary time between a trivial and a non-trivial SPT, and can therefore probe the edge physics.
%Correlators computed in this theory, of the type $\braket{\Psi_\text{Trivial}|\mathcal{O}_1 \mathcal{O}_2 |\Psi_\text{SPT}}$, measure observables on an edge in imaginary time between a trivial and a non-trivial SPT.
%$Z_\text{SPT}$ is the theory giving ``strange correlators'' \cite{Cenke2014},  $\braket{\Psi_\text{Trivial}|\mathcal{O}_1 \mathcal{O}_2 |\Psi_\text{SPT}}$, which measure observables on an edge in imaginary time between a trivial and a non-trivial SPT.
To complete the analogy with the FQH bulk-boundary correspondence, it was shown in \cite{PhysRevB.93.115105} that $\ket{\Psi_\text{SPT}}$ can be written in terms of correlators in the $Z_{\Psi_\text{SPT}}$ CFT.

%As shown in Ref.~\cite{PhysRevB.93.115105}, the bulk SPT wavefunction can be written in the charge basis as correlators in the $Z_{\Psi_\text{SPT}}$ CFT, in analogy with the FQH bulk-boundary correspondence.

%$Z_{\Psi_\text{SPT}}$ was already shown to map to an integrable loop model for the $SU(2)_1$ CFT
We will now calculate $Z_{\Psi}$ for both the gapped and gapless SPT.
Let us first briefly review the gapped case~\cite{Cenke2014,PhysRevB.93.115105}. We focus on the triangular lattice, but the results generalize straightforwardly to the Union Jack lattice.
		%Before moving to the gapless case, let us first identify the strange correlator theory of the gapped SPT as the calculation for gSPT will prove very similar.
%We will now identify the strange correlator theory for both the gapped and gapless SPT.
%We focus on the triangular lattice case for concreteness but there is a straightforward generalization to the Union Jack lattice.
%In the gapped case, 
	Starting from Eqs.~\eqref{eq:SPT_GS} and \eqref{eq:2DSPTPhase}, one can use the fact that $e^{i\theta_{\text{2D}}}$ factors over domain walls to analytically sum the $\tau$ degrees of freedom over each domain wall separately. This yields~\cite{PhysRevB.93.115105}
\begin{equation}
	Z_{\Psi_\text{SPT}} =  \sum_{\st{\sigma^z},\st{\tau^z}} e^{i\theta_\text{2D}(\sigma^z,\tau^z)} \propto \sum_{\{\text{dw}\}} n^{N[\text{dw}]} x^{L[\text{dw}]},
	\label{eq:SPT_SC}
\end{equation}
with $L[\text{dw}]$ the total length of domain walls, $N[\text{dw}]$ the total number of domain walls, $x^{-1}=\sqrt{2}$ and $n=2$.
Since domain walls form closed, non-intersecting loops, this can be identified as a dense (but not fully packed) loop model on the honeycomb lattice with loop fugacity $n=2$ and loop tension $x^{-1}=\sqrt{2}$ \cite{Jacobsen}.
%This can be identified as dense (but not fully packed) loop model with loop fugacity $n=2$ and loop tension $x^{-1}=\sqrt{2}$ \cite{Jacobsen}.
%where the sum runs over all configurations of domain walls of $\sigma^z$, $\st{\text{dw}}$, with total length $L[\text{dw}]$, and $N[\text{dw}]$ connected components.
%the sum over $\sigma^z$ has been replaced by a sum over domain wall configurations $\text{dw}$, $N[\text{dw}]$ is the number of domain walls and $L[\text{dw}]$ is the total length of all domain walls.
%Since domain walls form closed, non-intersecting loops, this is a dense (but not fully packed) loop model with loop fugacity $n=2$ and loop tension $x^{-1}=\sqrt{2}$ \cite{Jacobsen}.
%Using the fact that the phase $e^{i\theta_\text{2D}}$ factorizes over domain walls, the sum over $\tau$ degrees of freedom over each domain wall can be carried out analytically and one finds $n=2$ and $x^{-1}=\sqrt{2}$~\cite{PhysRevB.93.115105}.
For these parameters, this loop model is exactly solvable and is given by the $SU(2)_1$ CFT with central charge $c=1$ \cite{Jacobsen}, in agreement with the edge field theory given in Eq.~\eqref{eq:SPT_edge}.

%$\mathcal{L}$ is a \Romain{(dense but not fully packed)} loop configuration\Dan{, $N[\mathcal{L}]$ is the number of loops it contains and $L[\mathcal{L}]$ is its total length, }and with \Romain{loop weight} $n=2$ and \Romain{monomer fugacity} $x=1/\sqrt{2}$. For these parameters, this loop model is exactly solvable and is given by the $SU(2)_1$ CFT with central charge $c=1$.
%\Dan{Here we} used the fact that the phase factor factorizes over domain walls, and the sum over $\tau$ degrees of freedom over one domain wall can be carried out analytically \cite{PhysRevB.93.115105}.

The strange correlator theory for the gapless case can be calculated analogously by restricting to fully-packed loop configurations $\overline{\st{\text{dw}}}$.
%the only difference is that the sum over loops is now restricted to fully packed loops configurations, noted $\overline{\text{dw}}$.
%As a reminder, these fully packed loop configurations correspond to domain walls of maximally antiferromagnetic Ising spins on a triangular lattice.
%They can be obtained by taking the loop tension $x^{-1}$ to zero. 
%i.e. domain walls of a maximally antiferromagnetic spin configuration on the triangular lattice.
This leads to
\begin{equation}
	Z_{\Psi_\text{gSPT}} =  \sum_{\overline{\st{\sigma^z}},\st{\tau^z}} e^{i\theta_\text{2D}(\sigma^z,\tau^z)} \propto \sum_{\overline{\{\text{dw}\}}} n^{N[\overline{\text{dw}}]},
	\label{eq:gSPT_SC}
\end{equation}
%
%\begin{equation}
%	Z_{\Psi_\text{gSPT}} =  \sum_{\overline{\st{\sigma^z}},\st{\tau^z}}   \prod_{\text{dw}_{\sigma}} e^{i\theta_\text{1D}(\tau^z_{dw})} \propto \sum_{\overline{\mathcal{L}}} n^{N[\overline{\mathcal{L}}]},
%	\label{eq:gSPT_SC}
%\end{equation}
with $n=2$ again.
This loop model is also known to give a CFT, but with  $c=2$ instead~\cite{0305-4470-29-20-007}.
Hence the bulk wavefunction of gSPT can be written as a correlator in a $c=2$ CFT, which is good evidence for a $c=2$ edge.
It is remarkable that this imaginary-time edge picture holds for a non-relativistic bulk theory with $z=2$. Notice that our analysis has provided us with a natural way of interpolating from the gapped SPT to the gapless SPT by tuning the loop tension $x^{-1}$ from $\sqrt{2}$ to zero. 
In the following section, we will give further evidence by showing that the entanglement spectrum is given by a $c=2$ theory as well.

\begin{figure}
	\includegraphics[width=1.0\linewidth]{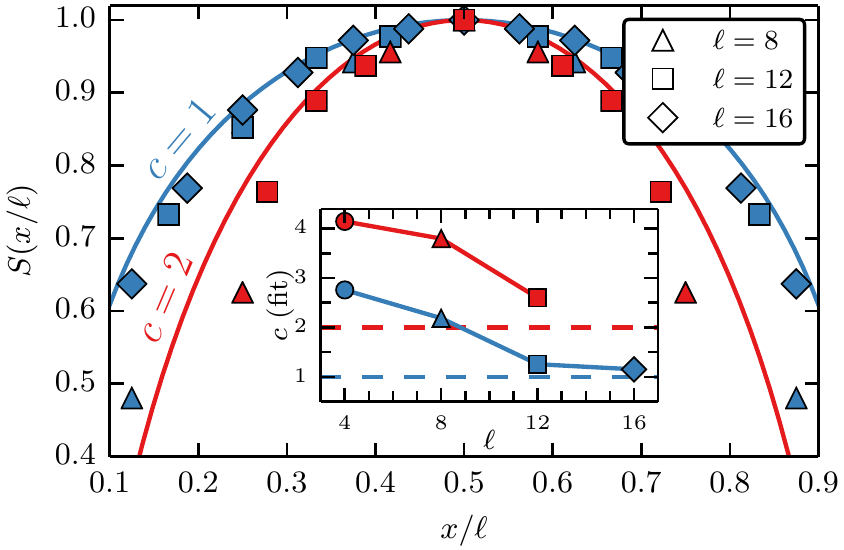}
	\caption{Entanglement entropy of the 1+1D ground state of the entanglement Hamiltonian $H_{\rm E} = -\log \rho$, with $\rho$ the reduced density matrix for a bipartite cut on an infinite cylinder of circumference $\ell$. Blue points are for $\ket{\Psi_\text{gTrivial}}$ and red points are for $\ket{\Psi_\text{gSPT}}$. Data is offset so that $S(1/2) =1$.}
		\label{fig:EE}
\end{figure}

\subsection{Entanglement Spectrum}

A useful property of entanglement cuts in systems that obey the area law for entanglement entropy \cite{RevModPhys.80.517,RevModPhys.82.277} is that the corresponding entanglement Hamiltonian can be interpreted as an ``edge'' Hamiltonian \cite{PhysRevLett.101.010504}.
This correspondence between entanglement and edge Hamiltonians has been shown rigorously in certain cases~\cite{PhysRevB.86.245310} and has been very useful in the numerical identification of various topological phases of matter.
While this technique has been mostly used so far for systems that are gapped in the bulk, we emphasize that this is not an inherent limitation.
As long as the area law is respected, it is always possible in practice to interpret the entanglement Hamiltonian as an edge Hamiltonian (see for example Ref.~\cite{PhysRevB.93.174414} where a gapless chiral spin liquid is shown to have an entanglement spectrum described by a CFT).

Consider $\ket{\Psi_\text{gTrivial}}$ and $\ket{\Psi_\text{gSPT}}$ on cylinders of circumference $\ell$ and infinite length. 
We make an entanglement cut transverse to the cylinder, which splits it into two semi-infinite regions, and we compute the reduced density matrix $\rho=e^{-H_E}$. As explained in the Appendix, using special properties of our exact ground state wavefunctions, it is possible to show that they satisfy the area law~\cite{PhysRevLett.97.050404}, to find the exact Schmidt decomposition, and thence compute the entanglement spectrum by numerical exact diagonalization of a two-dimensional transfer matrix. The reduced density matrix has support only on the entanglement cut, so it naturally describes a 1D system. Moreover, it has the form of a transfer matrix for a 2D statistical model, so the quantum-classical mapping provides $H_E$ as a local operator.

%
%
%We can apply a technique of Stephan \textit{et al.} to find the exact Schmidt decomposition. 
%The decomposition only depends on the states at the boundary and so the dimension of the reduced density matrix is proportional to $L$, the size of the cut. (We work on the square lattice for numerical convenience.) In an Appendix, we show that $\rho = S S^t$ where $S$ is the transfer matrix on the cylinder times a state-dependent normalization factor. 

We find that for both gTrivial and gSPT, the spectral gap of $H_E$ goes like $1/\ell$, indicating the edges are indeed gapless (see Appendix for details). 
To identify the CFT described by $H_E$ as a $1+1d$ theory on a circle, we compute the entanglement entropy $S(x)$ of the ground state of $H_E$ for cuts of length $x$.
We then apply the standard result of Cardy and Calabrese to extract the central charge~\cite{1742-5468-2004-06-P06002,1751-8121-42-50-504005}:
\begin{equation}
	S(x) = \frac{c}{3} \ln \sin \frac{\pi x}{\ell}.
	\label{eq:C_C_formula}
\end{equation}
%Let $\rho_0$ be the density matrix of the (1+1d) ground state of $H_E$. 
Figure \ref{fig:EE} shows $S(x)$ for the gTrivial and gSPT case. 
As  $\ell~\to~\infty$, this converges from below~\cite{zamolodchikov1986irreversibility} to \eqref{eq:C_C_formula}. 
%For $L$ spins on the $A$ lattice, the dimension of the Hilbert space is $2^{3L/2}$. In particular, this means $L=16$ for the gSPT state is dimension $2^{24}$, which is numerical impractical. 
%We restrict $L$ to multiples of 4 because multiples of 2 are in a different sector of the FPL theory
In the inset, the central charges are seen to converge to $c=1$ and $c=2$ for the gTrivial and gSPT orders respectively.

We may thus conclude, having shown it by three independent and consistent methods, that the gapless Trivial state has an edge mode with central charge $c=1$ while the gapless SPT case has $c=2$ --- recall however that only half of this edge is protected by the $\Z_2^3$ symmetry. These ballistic edge modes ($z=1$) in a diffusive ($z=2$) quantum critical system should have dramatic consequences for transport properties.

%\Dan{Mention the ``gauged'' version vaguely?}

\section{Conclusions}

Gapless symmetry protected order was proposed as a class of quantum matter. We provide a general construction for many gSPT systems by decorating the domain walls of gapless systems. To concretely understand gSPTs, we focused on two analytically solvable examples: a simple 1D system that extends the Ising model, and a gapless spin liquid in two dimensions. These demonstrate that gSPTs not only extend the crucial topological feature of SPTs --- robust gapless edge modes --- but also permit generalizations of tools developed for the gapped case, such as the bulk-boundary correspondence and the use of the entanglement spectrum as a probe of the edge. Both systems also exhibit exotic boundary behavior, including anomalous edge magnetization in the 1D example and, for 2D, $z=1$ edge dynamics for a $z=2$ system. Both in 1D and 2D, the gapless edge modes appear to be exponentially localized by the gap of the $\tau$ spins, even though they induce an algebraic disturbance for the $\sigma$ spins into the critical bulk.

 These are by no means the \textit{only} gSPTs. To wit, in the 1D example one could straightforwardly replace the Ising spins with parafermions or a Potts model; 2D should permit a gapless topological state with relativistic Majorana edge modes using Majorana chains as decoration~\cite{PhysRevB.94.115127,PhysRevB.94.115115}, and it might be possible to find 3D gapless spin liquids where analytic control over the decoration is possible~\cite{PhysRevB.69.064404,PhysRevB.91.195131,PhysRevX.6.011034}. Two-dimensional gSPT states could also be realized in realistic strongly correlated electronic systems~\cite{2015arXiv151101505S,PhysRevB.93.195141}.
 
 %Once a gapless system is identified, each choice of an SPT for decoration should provide a different gapless SPT. 
 %Our construction used different sublattices to impose a spatial distinction between the gapless system and its decoration, but this may be unnecessary. In a theory where a gapped and gapless part are not spatially distinct it may still be possible to form a gSPT by decorating the gapped sector. \Romain{Not sure I really like these two sentences }
 
More broadly, some of our examples can be interpreted as ``twisted'' quantum phase transitions between SPT and broken symmetry phases, which are expected to be more generic than direct transitions from trivial to SPT phases~\cite{2015arXiv151107460T,Tsui2015330,Lokman3}. 
Even if the bulk universality class of such twisted transitions is the same as for quantum critical points between trivial paramagnets and symmetry-broken phases (and hence described by conventional Ginzburg-Landau theory), our results indicate that twisted transitions differ from regular transitions in terms of surface criticality, in agreement with recent Monte Carlo results~\cite{PhysRevLett.118.087201}.
%
%\Thomas{Even if the bulk universality class of such ``twisted'' quantum critical points between SPT and symmetry-broken phases is described by a conventional Ginzburg-Landau theory, our results indicate that such transitions differ from regular transitions between trivial paramagnets and symmetry-broken phases in terms of surface criticality, in agreement with recent Monte Carlo results~\cite{PhysRevLett.118.087201}. }
From a field theory perspective, trivial paramagnets and gapped SPTs can be understood as non-linear sigma models in their gapped, disordered phase, the only difference being that the latter has a topological $\theta$ term with $\theta=2\pi$~\cite{PhysRevLett.61.1029,PhysRevB.91.134404}. It would be interesting to study the role of this $\theta$ term on the transition to symmetry-broken phases.

We also emphasize that our construction leads to gSPT states that are just as stable as the underlying gTrivial wavefunctions before applying the decoration. In particular, our construction yields stable gSPT phases by decorating Luttinger liquids in 1D (see section~\ref{secLLstar}) or $U(1)$ gauge theories in 3D (left for future work). It would also be interesting to relate gSPTs to other gapless topological states of matter, including gapless fractionalized states~\cite{PhysRevB.66.235110,2016arXiv160605652L,PhysRevB.89.174411}, in particular by partially gauging the symmetries~\cite{Levin2012}. We leave these directions for future work and we hope that gapless SPTs might provide a useful starting point to systematically study gapless topological matter.

\begin{acknowledgements}
We thank S. Parameswaran, A.C. Potter, Z. Ringel, S. Simon and B. Ware for insightful discussions, and Y.-M. Lu, A. Nahum, A.C. Potter, Z. Ringel and Y.-Z. You for useful comments on the manuscript. We acknowledge support from the Emergent Phenomena in Quantum Systems initiative of the Gordon and Betty Moore Foundation (T.S.), NSF DMR-1507141 (D.P.), and the Department of Energy through the Quantum Materials program of LBNL (R.V.).
\end{acknowledgements}

\bibliography{gapless_SPT}

%\clearpage

\appendix

\newcommand{\sdl}{\sigma_{\partial L}}
\newcommand{\sdr}{\sigma_{\partial R}}
\newcommand{\tdl}{\tau_{\partial L}}
\newcommand{\tdr}{\tau_{\partial R}}

\section{Entanglement Spectrum on the Cylinder}
\label{app_EE}
This Appendix computes the entanglement spectrum of the 2+1d gapless states introduced in Section IV.  Below we will explicitly calculate the reduced density matrix for both the ``gapless trivial'' and ``gapless SPT'' systems and show it may be written in terms of a transfer matrix for a gapless 1+1d system, which we interpret as the edge theory. Using techniques of 1+1d CFT \cite{1751-8121-42-50-504005,1742-5468-2004-06-P06002}, we demonstrate this edge theory has $c=1$ for the gapless trivial case but $c=2$ for the gapless SPT case.

\begin{figure}[t]	
	\includegraphics[width=3.3in]{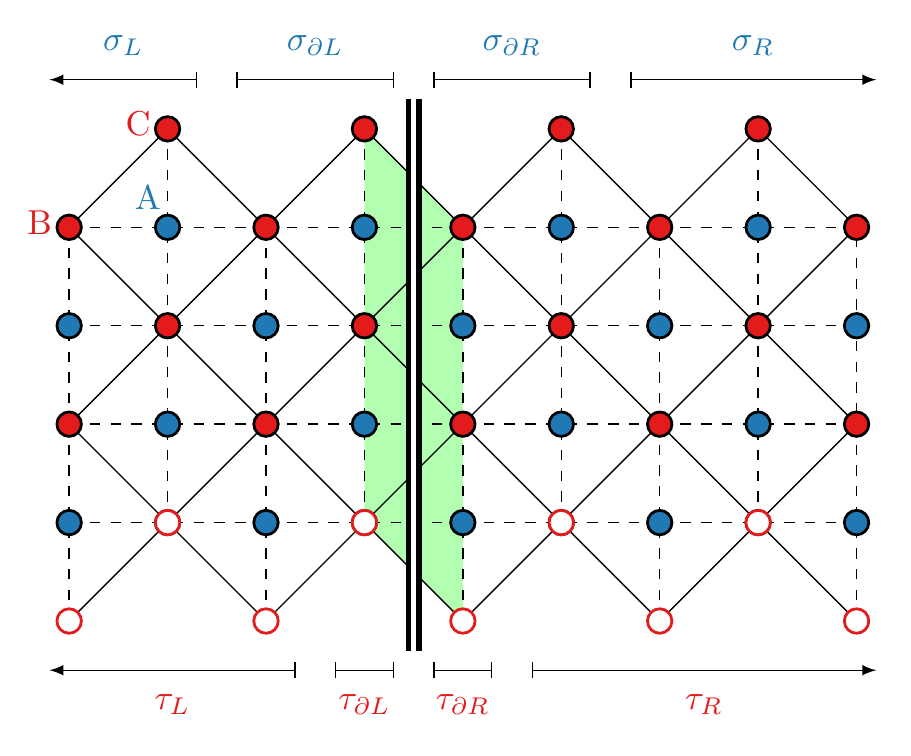}
	\caption{The Union Jack lattice showing the $A$ sublattice (blue) and BC sublattices (red) on a cylinder. The empty sites at the bottom are identified with the full ones on top, giving a circumference $\ell =4$. The entanglement cut is denoted with a double line, and the triangles ABC that it breaks are highlighted in green. The top and bottom show the extent of the left, right, and boundary regions for each species of spins.}
	\label{fig:entanglement_cut}
\end{figure}

Consider a cylinder with a circumference of $\ell$ and infinite length (see Figure \ref{fig:entanglement_cut}). 
The analytic results below are general for any geometry, but for numerical convenience, we work on the (tilted) Union Jack lattice. 
The circumference $\ell$ is defined so that every column is composed of $\ell$ sites.
Let us consider an entanglement cut transverse to the cylinder, which divides the cylinder into a left (L) and right (R) side. 
Fig. \ref{fig:entanglement_cut} shows the geometry and sets notation.

With notation from Figure~\ref{fig:entanglement_cut}, the gSPT wavefunction Eq. \eqref{eq:g_SPT_GS} can be written more explicitly as
\begin{align}
	\Ket{\Psi_\text{gSPT}} \ &=\ \frac{1}{\sqrt{Z}} \sum_{\overline{\st{\sigma}},\st{\tau}} e^{i \theta(L)} e^{i\theta(R)} e^{i\theta(\partial_L, \partial_R)}\\
	\label{eq:gSPT_wavefunction}
	\ &\hspace{2em}\times \ket{\sigma_L, \tau_L, \sdl, \tdl} \otimes \ket{\sigma_R, \tau_R, \sdr, \tdr},
	\nonumber
\end{align}
where the sum runs over all configurations of the $\sigma^z$ spins whose domain walls are fully-packed loops (FPL) and over all $\tau^z$ spins whatsoever. $Z$ factors into the partition function of the FPL model for $\sigma$ and a trivial normalization factor for $\tau$:
% for the FPL model times $2^{\# \tau}$ as a normalization factor with $\# \tau$ the number of $\tau$ spins:
\begin{equation}
\begin{aligned}
	Z=\sum_{\overline{\st{\sigma}},\st{\tau}} \ 1 = 2^{\# \tau} \sum_{\overline{\st{\sigma}}} \ 1,
	\label{eq:gSPT_wavefunction_Z}
\end{aligned}
\end{equation}
with ${\# \tau}$ the number of $\tau$ spins.
For a domain $D$, the phase factor $e^{i\theta(D)}$ gives a factor of $-1$ for each triangle strictly included in $D$ with three down spins.
%For a domain $D$, the phase factors $e^{i\theta(D)}$ gives a factor of $-1$ for each triangle in $D$ with three down spins and a factor of $+1$ otherwise. 
The triangles that cross the cut and contribute to $ e^{i\theta(\partial L, \partial R)}$ are highlighted in green in Fig.~\ref{fig:entanglement_cut}.

Define wavefunctions on the left side for each possible choice of spin configurations at the left boundary (denoted $\partial L$) by
\begin{equation}
	\begin{aligned}
	&\ket{\Psi_\text{gSPT}^L[\sdl, \tdl]}\\
	\ &=\ \frac{1}{\sqrt{Z_L[\sdl]}} \sum_{\overline{\st{\sigma_L}} ,\st{\tau_L}} e^{i \theta(L)} \ket{\sigma_L, \tau_L,\sdl,\tdl},
\end{aligned}
	\label{eq:left_wavefunction}
\end{equation}
where $Z_L$ is the partition function on the left side, and is independent of $\tau_L$. Define $\ket{\Psi_\text{SPT}^R[\sdr, \tdr]}$ analogously on the right side. In a dual picture, the domain walls of the $\sigma$ spins are isomorphic to configurations of the 6-vertex model. This local constraint would allow an exact Schmidt decomposition following~\cite{PhysRevB.80.184421}. However, different cuts make physical sense with domain walls instead of spins. (Indeed, using the domain walls leads to a factorization of the density matrix as a product of the $\sigma$ and $\tau$ degrees of freedom.) We emphasize, therefore, that one must work with the actual spins. Conveniently, one may still use the local constraint on the $\sigma$ spins together with the zero correlation length of the $\tau$ spins to find an exact Schmidt decomposition.

 We may rewrite the entire wavefunction as 
\begin{equation}
	\begin{aligned}
		&\Ket{\Psi_\text{gSPT}} \ =\ \sum_{\substack{\st{\sdl,\sdr}\\ \st{\tdl, \tdr}}}  \left(\frac{Z_L[\sdl] Z_R[\sdr]}{Z}\right)^{1/2} e^{i\theta(\partial L, \partial R)}\\
		\ &\hspace{2em} \times  T_{\sdl, \sdr} \; \ket{\Psi_\text{SPT}^L[\sdl, \tdl]} \otimes \ket{\Psi_\text{SPT}^R[\sdr, \tdr]},
\end{aligned}
	\label{eq:factored_wavefunction}
\end{equation}
where the sum over $\sdl$ and $\sdr$ is now \textit{unconstrained}. 
Here $T$ is the transfer matrix for the fully-packed loop model with loop fugacity one. 
Its role is to enforce the FPL constraint between the left and right sides. 
In the following, we will use the orthogonality property
\begin{equation}
	\braket{\Psi_\text{SPT}^L[\sdl', \tdl'] |\Psi_\text{SPT}^L[\sdl, \tdl]} = \delta_{\sdl',\sdl} \delta_{\tdl',\tdl}.
	\label{eq:schmidt_orthogonality}
\end{equation}
%Because
%\begin{equation}
%	\braket{\Psi_\text{SPT}^L[\sdl', \tdl'] |\Psi_\text{SPT}^L[\sdl, \tdl]} = \delta_{\sdl',\sdl} \delta_{\tdl',\tdl},
%	\label{eq:schmidt_orthogonality}
%\end{equation}
%and similarly for the right side, Eq.~\eqref{eq:factored_wavefunction} is a Schmidt decomposition. 
%This decomposition has the non-trivial property that only the spins on the boundary enter the sum in~\eqref{eq:factored_wavefunction}. 
%Generically, all the spins might have appeared.

Starting from the density matrix $\rho = \Ket{\Psi_\text{gSPT}} \Bra{\Psi_\text{gSPT}}$, we may use \eqref{eq:factored_wavefunction} to immediately write the \textit{reduced} density matrix on the left side:
\begin{equation}
	\braket{\Psi_\text{SPT}^L[\sdl, \tdl] |\rho_L| \Psi_\text{SPT}^L[\sdl', \tdl']} = \left( S S^t \right)_{\sdl\tdl,\sdl' \tdl'},
 \label{eq:reduced_density_matrix}
\end{equation}
where we used the above orthogonality property and where $S$ is a transfer matrix from the left to the right side
\begin{equation}
	\begin{aligned}
		&S_{\sdl,\tdl, \sdr, \tdr} = \left(\frac{Z_L[\sdl] Z_R[\sdr]}{Z} \right)^{1/2} \\
		\ &\hspace{2em}\times T_{\sdl\sdr} e^{i\theta(\sdl,\tdl,\sdr,\tdr)}.
\end{aligned}
	\label{eq:S_matrix}
\end{equation}

\begin{figure}[t!]	
	\includegraphics{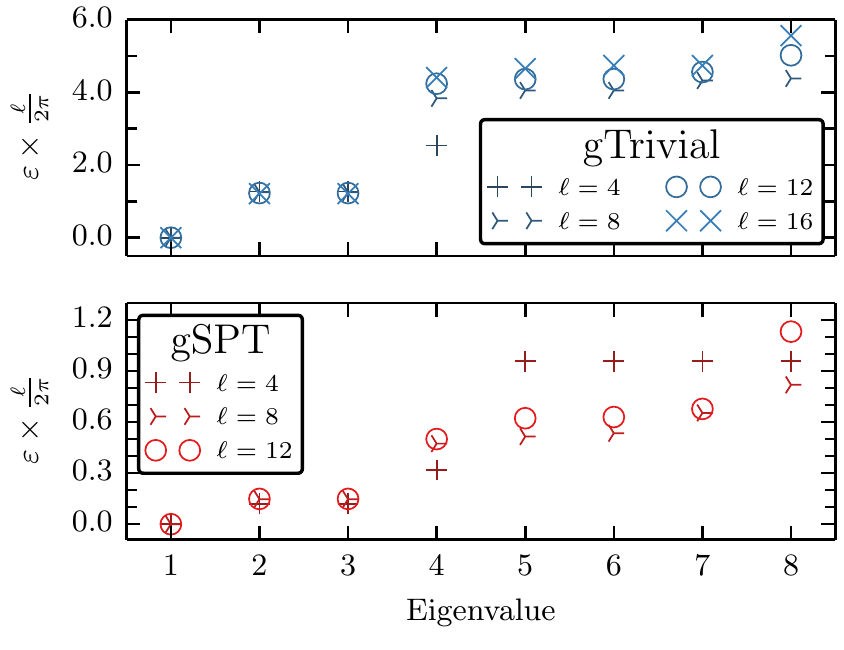}
	\caption{Entanglement spectral gaps for the gTrivial and gSPT for entanglement cuts of the 2+1d example on the cylinder of circumference $\ell$. The vertical axis is rescaled to be able to directly read off the operator dimensions of the excitations (up to a non-universal sound velocity). Only the first few excitations are well-converged for this range of $\ell$.}
		\label{fig:numerics}
\end{figure}

The reduced density matrix manifestly depends only on the degrees of freedom at the entanglement cut whereas generically it might have depended on all the spins on the left side. 
%On the infinite cylinder, this reduces $\rho_L$ from infinite to finite rank.	
If we define the entanglement Hamiltonian via $\rho_L = e^{-H_E}$, then $H_E$ describes a 1+1d system on the boundary degrees of freedom. To compute the spectrum of $H_E$ on the cylinder, we use the fact that
\begin{equation}
	\begin{aligned}
		Z_L[\sdl] \ &=\  \lim_{N \to \infty} 2^{\# \tau} \sum_{\st{\sigma}} \braket{\sigma|T^N|\sdl}\\
		\ &=\    \lim_{N \to \infty} 2^{\# \tau} \sum_{\st{\sigma}} \braket{\sigma|R} \lambda^N \braket{L|\sdl}, 
\end{aligned}
\end{equation}
where by the Perron-Frobenius theorem, $T^N \rightarrow \ket{R} \lambda^N \bra{L}$ where $\lambda$ is the largest eigenvalue of $T$ and $\ket{R}$ and $\ket{L}$ are the corresponding right- and left-eigenvectors.
The sum runs over all configurations of $\sigma^z$ on one column.
% Because $T$ is non-symmetric, there is not necessarily a unique largest eigenvalue, and indeed this is only the case when $\ell$ (the number of $\sigma$ sites on the boundary) is a multiple of four, so we restrict to that case. Other even $\ell$'s correspond to a different sector of the FPL CFT. 
 This implies
\begin{equation}
	\begin{aligned}
		&S_{\sdl,\tdl, \sdr, \tdr}\\
		\ &=\ \left(\frac{\braket{L|\sdl} \braket{\sdr|R}}{\braket{L|R} \lambda\;  2^{\ell}}\right)^{1/2} T_{\sdl\sdr}  e^{i\theta(\sdl,\tdl,\sdr,\tdr)}.
\end{aligned}
	\label{eq:S_matrix_computable}
\end{equation}
One can check that this is properly normalized: $\operatorname{Tr} \rho_L = \operatorname{Tr} S S^t = 1$.

%Because $T$ has values of either one or zero and $\left( e^{i\theta} \right)^2 \equiv 1$, this is properly normalized: $\operatorname{Tr} \rho_L = \operatorname{Tr} S S^t = 1$.

We now employ exact diagonalization. At size $\ell$, $S$ is a $2^{3\ell/2} \times 2^{3\ell/2}$ matrix, making exact diagonalization practical for $\ell = 4,8,12$. 
We restrict to $\ell$ being a multiple of four in order to stay in the symmetric ground state sector of the loop model.
Since the $\tau$ part of the matrix is dense, larger sizes are impractical. However, in the gapless trivial case, we may discard the $\tau$ part and work on larger systems. For both the gapless trivial (where $e^{i\theta} \equiv 1$) and gapless SPT cases, the spectral gap for $H_E$ goes as $1/\ell$, which indicates gaplessness with dynamical exponent $z=1$. This is shown in Fig.~\ref{fig:numerics}.

By looking at the ground state of $H_E\ket{0} = \varepsilon_0 \ket{0}$, we may determine the central charge of $H_E$ by making entanglement cuts in the (1+1d) edge system and comparing to the Cardy-Calabrese equation \eqref{eq:C_C_formula}. For each $\ell$, a one parameter fit to the Cardy-Calabrese result was performed to extract the central charge. Fig.~\ref{fig:EE} shows that the central charge converges to $c=1$ in the gapless trivial case and $c=2$ in the gapless SPT case.

\end{document}